\documentclass[12pt, a4paper]{article}

\usepackage[T1]{fontenc}

\usepackage{authblk}
\usepackage{graphicx}
\usepackage{subcaption}
\usepackage[title]{appendix}
\usepackage{amsmath}
\usepackage{lineno}
\usepackage{float}
\usepackage[round]{natbib}
\usepackage[none]{hyphenat}
\usepackage[a4paper, portrait, margin=1in]{geometry}
\usepackage{color}
\usepackage{array}
\usepackage[flushleft]{threeparttable}
\usepackage{setspace}
\usepackage[skip=0.5pt]{caption}
\usepackage[bottom, hang, flushmargin]{footmisc}
\usepackage{pdflscape}

\usepackage[hyphens]{url}

\newcolumntype{L}[1]{>{\raggedright\let\newline\\\arraybackslash\hspace{0pt}}m{#1}}
\newcolumntype{C}[1]{>{\centering\let\newline\\\arraybackslash\hspace{0pt}}m{#1}}
\newcolumntype{R}[1]{>{\raggedleft\let\newline\\\arraybackslash\hspace{0pt}}m{#1}}

\newcommand\blfootnote[1]{%
  \begingroup
  \renewcommand\thefootnote{}\footnote{#1}%
  \addtocounter{footnote}{-1}%
  \endgroup
}

\usepackage{multirow}
\begin{document}
\doublespacing
\onehalfspacing
\title{The English Patient: Evaluating Local Lockdowns Using Real-Time COVID-19 \& Consumption Data}
\singlespacing

\author{John Gathergood\footnote{Department of Economics, Nottingham University. john.gathergood@nottingham.ac.uk} \\ Benedict Guttman-Kenney\footnote{Corresponding author. Chicago Booth School of Business, University of Chicago. benedict@chicagobooth.edu}}

\date{January 11, 2021}

\maketitle

\begin{abstract}

\noindent
We find UK ``local lockdowns'' of cities and small regions, focused on limiting how many people a household can interact with and in what settings, are effective in turning the tide on rising positive COVID-19 cases. Yet, by focusing on household mixing within the home, these local lockdowns have not inflicted the large declines in consumption observed in March 2020 when the first virus wave and first national lockdown occurred. Our study harnesses a new source of real-time, transaction-level consumption data that we show to be highly correlated with official statistics. The effectiveness of local lockdowns are evaluated applying a difference-in-difference approach which exploits nearby localities not subject to local lockdowns as comparison groups. Our findings indicate that policymakers may be able to contain virus outbreaks \textit{without} killing local economies. However, the ultimate effectiveness of local lockdowns is expected to be highly dependent on co-ordination between regions and an effective system of testing. \vfill



\end{abstract}

\tiny
\blfootnote{
First version 8 October 2020 with updated drafting and references. The views expressed are the authors and do not necessarily reflect the views of Fable Data Limited.
We thank Fable Data Limited for sharing these data for research.
Thanks to Constantine Yannelis, Chad Syverson, Kilian Huber, Pietro Veronesi, Scott Nelson, Pascal Noel, Peter Ganong, Jack Light, Hans-Joachim Voth, one anonymous referee and Chicago Booth Finance Student Brownbag audience for their feedback.
We are grateful to Suraj Gohil, Debbie Mulloy, Fiona Isaac, Zina Papageorgiou and Sairam Kamath at Fable Data Limited and Lindsey Melynk and Rich Cortez at Chicago Booth for their help facilitating this research. This work is supported by the UK Economic and Social Research Council (ESRC) under grant number ES/V004867/1 `Real-time evaluation of the effects of Covid-19 and policy responses on consumer and small business finances'.}

\normalsize
\doublespacing

\thispagestyle{empty}\clearpage
\section{Introduction}
How can COVID-19 cases be contained without causing damage to the economy? This question dominates the thinking of policymakers, who face a seemingly uncomfortable trade-off between limiting virus transmission in the economy via reducing social contact, and maintaining economic activity which relies on social contact for the production and consumption of goods and services. 

The first wave of COVID-19 in early 2020 saw most nations adopt stringent restrictions on social contact in almost all settings in order to contain the spread of the virus.
These restrictions severely hindered the means of production and consumption in the economy, leading to large drops in output from a combination of the virus and such restrictions.
However, recent improvements in testing and tracing leading to identification of clusters of cases in high-infection areas have facilitated a more targeted, localised approach to applying restrictions to social contact, known as ``local lockdowns''.
This approach to limiting the spread of COVID-19 was identified early-on in the pandemic as a beneficial strategy.\footnote{See \url{https://medium.com/@tomaspueyo/coronavirus-the-hammer-and-the-dance-be9337092b56}}
In some countries -- such as in the UK -- governments have legal powers to implement such local measures, but this has been a source of political tension between local and national authorities.\footnote{The legal power of the government in the UK to impose local restrictions contrasts with other nations, such as the US and Spain, where such measures can only be implemented by local governments. Nonetheless, both approaches have resulted in local resistance to such measures due to concerns of the adverse effects on local economies -- with a standoff between local and national governments over locking down Madrid, anti-lockdown protests in London, Van Morrison releasing protest songs over the Northern Ireland local lockdown and some mayors and local governments demanding responsibilities powers and associated funding be delegated to them.}
The centralized UK approach offers a particularly interesting contrast to the US, where the policy response has limited national co-ordination -- and none in regards to lockdowns.

In this paper, we are the first academics to use a new source of real-time and highly granular European consumption data.
We combine these with data on coronavirus cases to analyse the impact of local lockdowns on both COVID-19 cases and local consumption.
Using a difference-in-difference methodology, we estimate the impact of local lockdowns imposed in the late summer of 2020 on a series of UK cities, examining their ability to contain of COVID-19 cases and how consumer spending responded.

Our study makes two main contributions.
First, we introduce a new source of real-time, transaction-level consumption data -- Fable Data -- that can be used for economic research and to inform policymaking.
These data contain transaction-by-transaction spending data, updated daily, for large representative samples of UK bank accounts and credit cards, with individual-level identifiers and geocode identifiers.
We show these data are a highly correlated, leading indicator of official Bank of England statistics -- data that are only available in aggregated form and with many months lag -- in contrast to Fable Data which are available in real-time and disaggregated (correlation coefficient of 0.91 January 2018 - June 2020).
These data are applicable to a broad variety of questions in the analysis of individual consumption behavior.
They present a new opportunity for researchers to measure consumption in complementary and arguably more reliable ways than using data from consumption surveys, which has become less reliable in recent decades and has prompted a variety of initiatives aimed at improving the measurement of consumption \citep[see][]{browning2014measurement, landais2020introduction}.
These data show the UK's economic recovery in spending April to August 2020 had stalled in September and October.

Second, we advance understanding of the economic costs of mitigation strategies to contain the second wave of COVID-19.
This user-case is economically important and policy-relevant to the time of writing, as policymakers around the world are grappling with how to both contain second or third waves of virus outbreaks and also to keep economic recoveries going.
Local lockdowns are also a source of tensions between national and local governments and thus our research may help to inform such disputes. 
The UK local lockdowns we study apply to cities or small regions.
They restrict how many individuals people can mix with, in what settings (e.g. restaurants) and under what requirements (e.g. outside, wearing masks).
A key policy design was trying to enable people to keep consuming in COVID-19 secure settings while limiting their interactions in less secure settings (e.g. households visiting each others' homes).
We show that UK local lockdowns to contain virus outbreaks -- covering one in four people by September 2020 -- typically turn the tide on rising COVID-19 cases though there is heterogeneity in such results.\footnote{\url{https://www.bbc.com/news/uk-england-52934822}}

We do not find evidence such local lockdowns resulted in large spending declines: observing little, if any, declines.
We use a difference-in-difference design that compares the evolution of daily consumption in an area subject to a local lockdown compared to a similar, nearby locality not subject to such restrictions.
By using daily data we can precisely estimate our results relative to the timing of local lockdowns being announced.
Our difference-in-difference approach is designed to be interpreted as descriptive not causal.\footnote{We typically observe common pre-trends between control and treatment groups, however, we do observe noticeable increases in the number of positive COVID-19 cases for the treatment groups just before and after local lockdowns. The nature of local lockdowns explain this behavior -- a key component is to increase testing capacity and thus the number of positive cases will be expected to rise. However, some areas subject to local lockdowns had rises that appear too large and sharp to be driven by differential testing and in such cases the control localities are less suitable counterfactuals.}
These data show -- both for treatment and control localities -- large drops in consumption when the March 2020 first virus wave and national lockdown occurred.
We conclude there are little, if any, declines in spending from the local lockdowns: certainly not of the magnitude of the March decline.
Estimates for the time-path of cases, in contrast, show that while COVID-19 cases typically continue to rise following the onset of a local lockdown (as measures take time to have effect) they then start to stabilize: indicating the local lockdowns had some short-term success.

While our evidence indicates some initial successes from local lockdowns, in late September and October 2020 the UK (along with many European countries) experienced a rapid, nationwide rise in COVID-19 cases (the cause of which is not yet clear). This has led to more restrictive regional measures being introduced in mid-October including shutting down businesses and before second and third national lockdowns were imposed in November 2020 and January 2021.
The COVID-19 positive case rates accompanying these more recent lockdowns were far higher than the rates in the local lockdowns we examine in this paper.

Our study contributes to a burgeoning literature understanding the economic effects of COVID-19.
A variety of early studies showed how the onset of COVID-19 dramatically changed consumption behavior.
The first study to do so was \cite{baker2020does} using US fintech data and following this Opportunity Insights \citep{chetty2020did,chetty2020real} produced a dashboard using multiple data sources to track regional US consumption behavior alongside other economic indicators.\footnote{https://tracktherecovery.org}
Beyond the US similar exercises have been carried out to understand household consumption in the early stages of the pandemic -- showing remarkably consistent results \citep{andersen2020consumer,bounie2020consumers,ifs2020,campos2020consumption,carvalho2020tracking,chen2020impact,chronopoulos2020consumer,horvath2020covid,surico2020consumption,watanabe2020online}.
Analysis of JP Morgan Chase data \citep{cox2020initial,farrell2020consumption} has described in detail how household balance sheets have changed as a result of the COVID-19 recession and how households have responded to fiscal stimulus.
A variety of studies have examined the effects of the first set of lockdowns on economic behavior and evaluating the degree to which there are trade-offs between policy interventions attempting to contain the virus and economic damage \citep{aum2020covid,beach20201918,barro2020coronavirus,coibion2020cost,correia1918pandemics,cui2020covid,dave2020were,friedson2020did,hacioglu2020distributional, glover2020health,goolsbee2020covid,goolsbee2020fear,guerrieri2020macroeconomic,hall2020trading,lilley2020public,miles2020living,jones2020optimal,toxvaerd2020equilibrium,wang2020covid}.

\section{Data}

\subsection{Consumption Data}

We combine data on cases of COVID-19 identified by the UK's testing framework with consumption data provided by Fable Data Limited.\footnote{Daily COVID-19 case data by Local Authority District is available at \url{https://coronavirus.data.gov.uk/}. More information on Fable Data is available at \url{www.fabledata.com}.} Fable data record hundreds of millions of transactions on consumer and SME spending across Europe from 2016 onwards.\footnote{Commercial sensitivities mean we do not disclose the exact number of accounts and transactions available in the data.}
Fable's transaction data are anonymized and available in real-time: our research access is with a one working day lag.
Fable sources data from a variety of banks and credit card companies: accounts cover both spending on credit cards and inflows and outflows on current (checking) accounts.
Data is at the account-level and hence we can follow spending behavior on an individual account over time.\footnote{In cases where one individual has multiple accounts, we cannot link multiple accounts in the data to the individual but can aggregate to a geographic region.} Fable data is similar to recently- available data sets from financial aggregators and service providers, but does not have some of the limitations of other datasets and uses anonymised customer data.\footnote{\cite{baker2018debt} provides validation and application of US financial aggregator data. Financial aggregator data for the UK is widely shared for research purposes by MoneyDashBoard, UK-based a fintech \citep{chronopoulos2020consumer,ifs2020,surico2020consumption}. \cite{ifs2020} analyse the characteristics of MoneyDashBoard users.}

For each spending transaction we observe a standard classification merchant category code for the spending type.  Fable also produces its own categorizations of spending, utilizing the more granular information it has available from transaction strings. These data also differentiate between online and store-based transactions.

For each UK account we observe the postcode sector of the cardholder's address.
In the UK, postcode sectors are very granular geographies: There are over 11,000 postcode sectors in the UK with each sector containing approximately 3,000 addresses.
Where a transaction can be linked to a particular store, the full address of that store is available.
Where a transaction is of a listed firm, Fable tags merchants to their parent groups and stock market tickers.

For this study we focus on transactions denominated in British pounds sterling on UK-based credit card accounts held by consumers.\footnote{We drop 113 individual credit card transactions over \pounds 50k as such outliers are unlikely to be consumer transactions and may distort results for very small geographic regions.}
The median and mean transaction values are £15 and £39 respectively.

Transaction-level spending data is highly volatile -- even with such large volumes of transactions -- and we observe strong movements at high frequency due to seasonality and day of week effects.
We therefore follow an approach to smooth the transaction volumes over time as used by Opportunity Insights on similar US data \citep{chetty2020real,chetty2020did}: aggregating spending by day at the level of geography of interest, taking a seven day moving average and dividing by the previous year's value.\footnote{For 29 February 2020 we divide by an average of 28 February and 1 March 2019.}
Finally, we normalize the series setting an index to 1 using the mean value 8 - 28 January 2020. We also construct daily series using a 14 and 28 day moving averages in a analogous fashion.

\subsection{Comparison with Official Statistics}

Fable data have many useful features, such as timeliness (it is available the next working day whereas official statistics are typically available only with a lag of several months), geographic granularity (being available at a lower level than official statistics) and, transaction- level (enabling more flexible analysis than aggregated official statistics).
These data can therefore potentially be used to construct leading indicators for policymakers and enable researchers to answer a broader set of research questions than was previously possible using prior data sources.

However, while these features are potentially valuable, their usefulness depends in part on how this data series relates to comprehensive, official data. To explore this, Figure \ref{fig:boe}, Panel A compares the time series of Fable Data UK annual changes in monthly credit card spending to the Bank of England series and shows they are highly correlated: correlations 0.91 (January 2018 to July 2020), 0.90 (January 2019 to July 2020) and 0.98 (January 2020 to July 2020).
Bank of England data is only published in aggregated form monthly and with a lag (e.g. July's data was published at the start of September).

Figure \ref{fig:boe}, Panel B shows Fable data measures for 7, 14, 28 day moving averages -- which can be calculated daily in real-time -- compared to the monthly series (which requires waiting until month end).
These daily moving averages show the sharp drop in consumption in March 2020 far earlier than the monthly series.
We thus conclude that we can use these data as a reliable real-time predictor of official data and as a reasonable proxy for measuring credit card spending.

On aggregate we observe the sharp fall in UK credit card spending near the time of the spike in Covid-19 cases and national lockdown announcement on 23 March 2020 and then a fairly steady recovery May - August.
Components of the national lockdown were ended in June and July but we do not observe rapid boost after these was lifted -- indicating spending may have been suppressed by fear of the virus during this early phase of the pandemic.
In September and October 2020, we observe these data to show that the recovery in spending since April 2020 has stalled and has started to decline.

In a companion paper \citep{gathergood2020lev} we use these same data to show the heterogeneous impacts of the COVID-19 crisis across UK regions and urban geographies: splits that are not possible using official statistics.\footnote{These figures were originally in an online appendix to the first version of this paper.}

\section{Local Lockdowns}

We use public data released by the UK government which details the areas affected by local lockdowns, including the dates of introduction and cessation of lockdown measures.
Each week Public Health England publishes a COVID-19 Surveillance Report that includes `The Watchlist' showing the incidence (and trend) of COVID-19 in local government areas (lower tier local authorities), whether household mixing is prohibited and lists areas on the watchlist.\footnote{\url{www.gov.uk/government/publications/national-covid-19-surveillance-reports}}
Scotland, Northern Ireland and Wales have comparable data that we gather.\footnote{\url{https://coronavirus.data.gov.uk/about-data#cases-by-lower-super-output-area-lsoa}, \url{www.opendata.nhs.scot/dataset/covid-19-in-scotland/resource/e8454cf0-1152-4bcb-b9da-4343f625dfef},\url{www.health-ni.gov.uk/publications/daily-dashboard-updates-covid-19-september-2020},\url{gov.wales/testing-data-coronavirus-covid-19-13-september-2020}}

Areas are added to the Watchlist considering a variety of metrics using professional judgment of UK public health officials according to the UK Government's COVID-19 Contain Framework'.\footnote{\url{www.gov.uk/government/publications/containing-and-managing-local-coronavirus-covid-19-outbreaks/covid-19-contain-framework-a-guide-for-local-decision-makers}}
Areas on the Watchlist fit into one of three categories:
\begin{itemize}
    \item \textbf{Concern Areas} - Local area is taking targeted actions to reduce COVID-19's prevalence (e.g. additional testing in care homes and increased community engagement with high risk groups).
    
    \item \textbf{Enhanced Support Areas} - More detailed plan agreed with the national team and with additional resources being provided to support the local team to control COVID-19 (e.g. epidemiological expertise, additional mobile testing capacity).
    
    \item \textbf{Intervention Areas (`Local Lockdowns')} - Divergence from the measures in place in the rest of England because of the significance of COVID-19's spread, with a detailed action plan in place, and local resources augmented with a national support to control COVID-19.
\end{itemize}

For this research we focus on local lockdowns.
By September 2020, approximately one in four people in the UK were subject to a local lockdown.
A key feature of such lockdowns is imposing restrictions on household mixing (e.g. preventing a tea party in someone's house) but permitting visits to more COVID-19 secure settings (e.g. having tea outdoor at a restaurant with strict hygiene and social distancing standards) in order to encourage consumers to keep spending while also trying to contain the virus.
Across the local lockdowns there was variation in the how much household mixing was restricted (e.g. in Caerphilly residents were not supposed to leave nor new people come in).\footnote{\url{https://www.bbc.com/news/uk-england-52934822}}
This contrasts with the March 2020 lockdown where there was a stay at home order not only preventing household mixing but also closing all non-essential shops and banning public activities (e.g. sports).

There were also nationwide (including in areas subject to local lockdowns) government financial incentives to encourage consumers to spend for much of the period of time we study. The most notable of these were cuts to sales taxes (Value Added Tax, VAT) on food, accommodation and attractions from 20\% to 5\% from July 2020 until January 2021 and `Eat Out to Help Out' scheme offering a 50\% discount (up to £10 per person on food and non-alcoholic drink) for eating out Mondays, Tuesdays and Wednesdays during August 2020.
 
\section{Methodology}

We use a difference-in-difference methodology to estimate the relationship between local lockdowns and daily consumption.
The use and challenges of difference-in-differences to estimate causal effects in COVID-19 is summarized in \cite{goodman2020using}. 
Our estimates provide a description of the evolution of COVID-19 cases and consumer spending pre- and post-lockdown on lockdown affected cities and comparison areas.
We do not interpret our estimates as showing causal relationships - our comparison cities are not perfect counterfactuals for the evolution of COVID-19 cases or consumer spending in the absence of a lockdown. 

We isolate the date local lockdowns were announced for each locality.
There was great uncertainty in both where and when such lockdowns would be introduced and the precise restrictions that they would require - as  reported in news reports at the time.
Some anticipatory behavior is possible given the data on covid cases was regularly published and reported on but as the threshold for intervention was low areas often suddenly appeared subject to local lockdowns - most notably in the case of Manchester where the mayor and city council were taken by surprise.\footnote{\url{https://www.bbc.com/news/uk-england-manchester-53592240}}
Table \ref{fig:dates} lists the timing of local lockdown announcements, the local authorities affected and the control group localities to compare against.
Where there are multiple localities in the same area subject to a lockdown announcement on the same day we aggregate data (e.g. South and North Lankarkshire to Lanarkshire) into a single `local authority group'.
For some areas subject to local lockdowns (e.g. Belfast) no suitable control group city exists.

We display descriptive results for thirteen pairs of treatment and control localities -- but primarily focus on the Manchester lockdown as that has a large sample, Liverpool offers a good control, the announcement was sudden and unexpected (being announced 9.15pm on a Thursday to coming into effect at midnight) and the case is particularly informative for considering the effects of a London lockdown.\footnote{\url{https://www.bbc.com/news/uk-england-manchester-53592240}}
The other areas studied that were subject to local lockdowns are: Aberdeen, Birmingham, Bolton, Caerphilly, Glasgow, Greater Glasgow, Lanarkshire, Leeds, Leicester, Newcastle, Preston and Wolverhampton.
For consumption measures, we allocate spending based on cardholder's address except for considering large store chains where we estimate results based on store location.
For COVID-19 cases allocation is as provided based on case reporting localities.

We first describe the time series for spending in these regions.
A standard differences- in-differences approach uses a parsimonious regression estimation approach, such as that presented in (Equation \ref{eq1}). The outcome of interest $Y_{g,t}$ is offline credit card spending measured as changes in an index of yearly changes in 7 day moving average of spending. The unit of observation is a day ($t$) for each local authority group ($g$) either subject to the local lockdown or the control group to compare it to.

\begin{equation}
Y_{g,t} = \alpha + \beta \; Treat_{g} + \delta_t \; Treat_{g} \; * \;  After_{t} + \gamma_t + \varepsilon_{g,t}
\label{eq1}
\end{equation}

To explain our methodology we draw upon the example of Manchester.
Manchester is the treated group ($Treat_g=1$) subject to the local lockdown and we use a `similar area' -- Liverpool -- as a control group ($Treat_g=0$).
These areas were chosen as cities of comparable size, in the same part of the country and showing similar pre-trends.
$After_t$ is an indicator equal to one if the time period is after the local lockdown announcement.
$(Treat_{g} \; * \;  After_{t})$ is the interaction of the above two terms: it is an indicator equal to one if the time period is after the local lockdown announcement and the area is in the treated group.
The difference-in-difference estimation approach allows for these areas to have different time-invariant relationships with consumption ($\alpha$ for Liverpool and $\alpha+\beta$ for Manchester) and $\gamma_t$ is a series of daily dummies (with $t=-1$ omitted) to control for any common time-varying factors (e.g. national changes in COVID-19 cases, economic policies).
$\delta_t$ provides an estimate for the relationship between local lockdown and consumption.

To better understand the local lockdowns we modify Equation \ref{eq1} to estimate a dynamic specification creating weekly dummies ($After_{W,g}$) for the weeks preceding ($W \in \{-3,-2,$ $-1,0\}$) and following  ($W \in \{1,2,3,4\}$) the lockdown announcement where the omitted weekly dummy ($W=0$) is the seven days preceding the local lockdown announcement ($t=-7$ to $t=-1$) and we use data up to four weeks pre- and post-lockdown (where available).
We focus on a short time window given the volatile period we study the control groups are likely to become less suitable comparisons over time - including the subsequent government introduction of a system of tiered lockdowns in the autumn meaning no untreated control group remained.
We estimate this using an OLS regression weighting observations by their 2019 resident population and cluster standard errors by region (with one observation per treatment group per day).\footnote{Resident population estimates are from Office for National Statistics (ONS).}

\begin{equation}
Y_{g,t} = \alpha + \beta \; Treat_{g} + \sum_{W \not = 0} \delta_k \; ( Treat_{g} \; * \;  After_{W,g,t} ) + \gamma_t + \varepsilon_{g,t}
\label{eq2}
\end{equation}

We regard our estimation approach as providing informative, descriptive real-time evidence to inform policymakers.
For these to be interpreted as causal effects for the effect of local lockdown on local spending would require a `common trends' assumption that, for example, in the absence the Manchester local lockdown, consumption in Manchester would have followed the common trend to that in Liverpool.
Such an assumption is unlikely to hold in these data - particularly over longer time horizons there may be spillovers between areas and other government measures being introduced - hence we interpret our short-term results as helpful descriptive evidence which might strongly suggest a causal relationship, due to what is known more generally about virus transmission and the efficacy of social distancing measures.

\section{Results}

\subsection{COVID-19 Cases}

\begin{figure}[H]
	\caption{\textbf{COVID-19 Cases in Lockdown Cities (yellow) vs Comparison Cities (black)}} 
	\centering
	\begin{tabular}{c} \\
		\textbf{Manchester} \\ 
		{\includegraphics[height=3in]{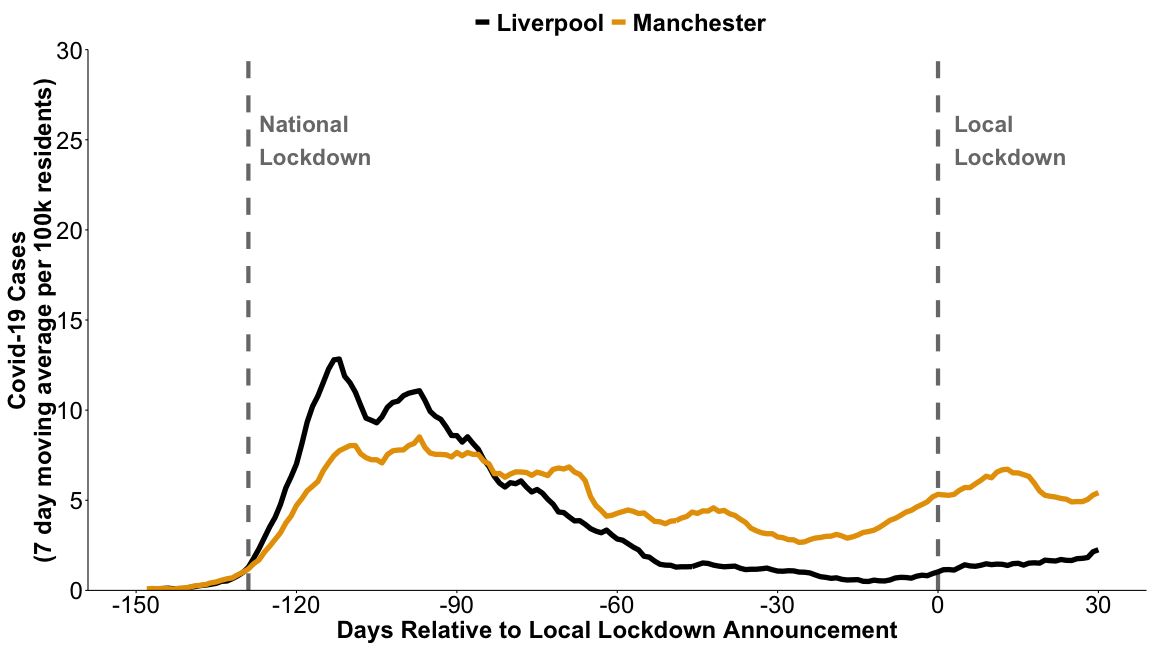}} \\
	\end{tabular}
	\begin{tabular}{c c c} \\
		\textbf{A. Aberdeen} &
		\textbf{B. Birmingham} &
		\textbf{C. Bolton} \\
		{\includegraphics[height=1in]{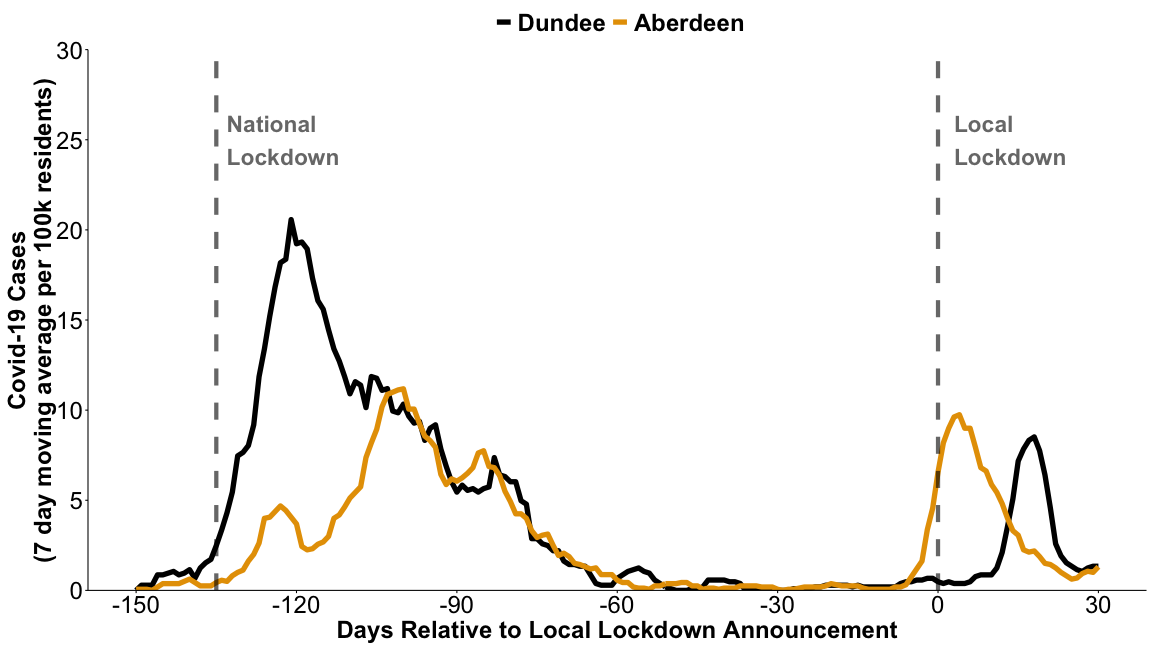}} & {\includegraphics[height=1in]{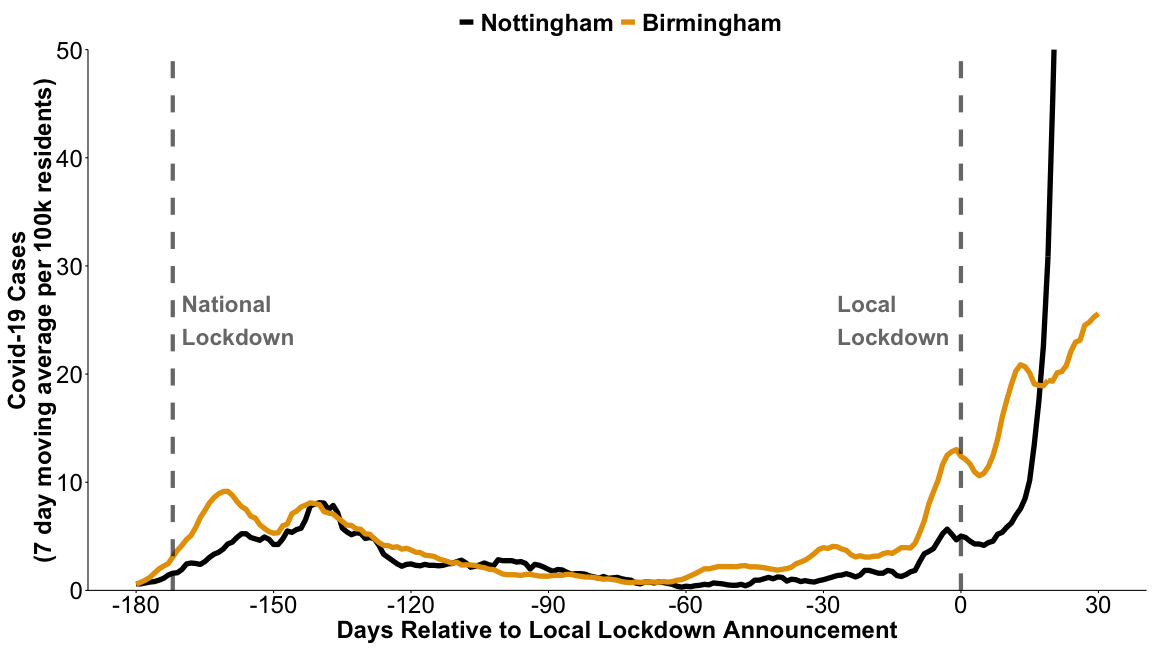}} & {\includegraphics[height=1in]{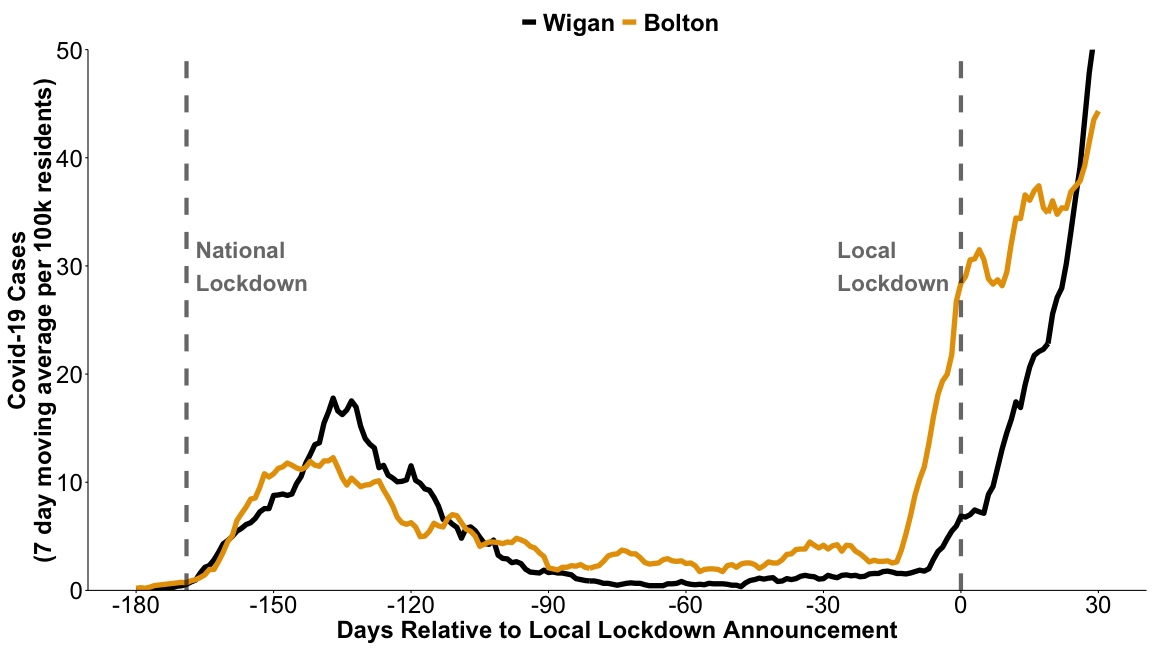}} \\ 
		\textbf{D. Caerphilly} &
		\textbf{E. Glasgow} &
		\textbf{F. Greater Glasgow} \\
		{\includegraphics[height=1in]{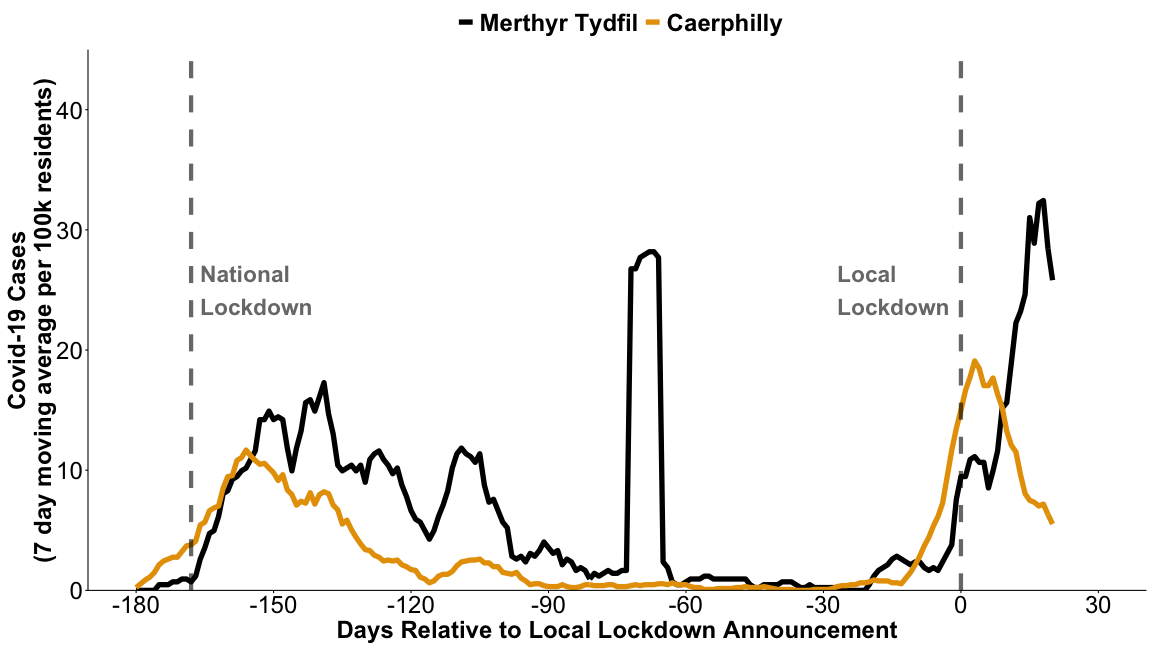}} & {\includegraphics[height=1in]{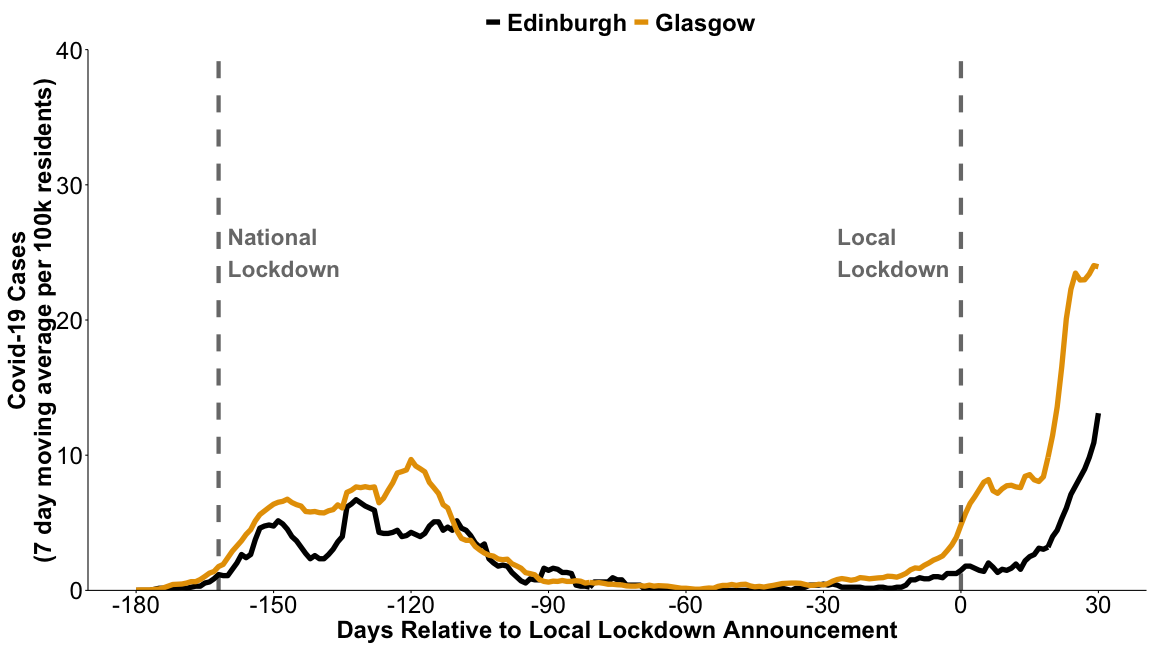}} & {\includegraphics[height=1in]{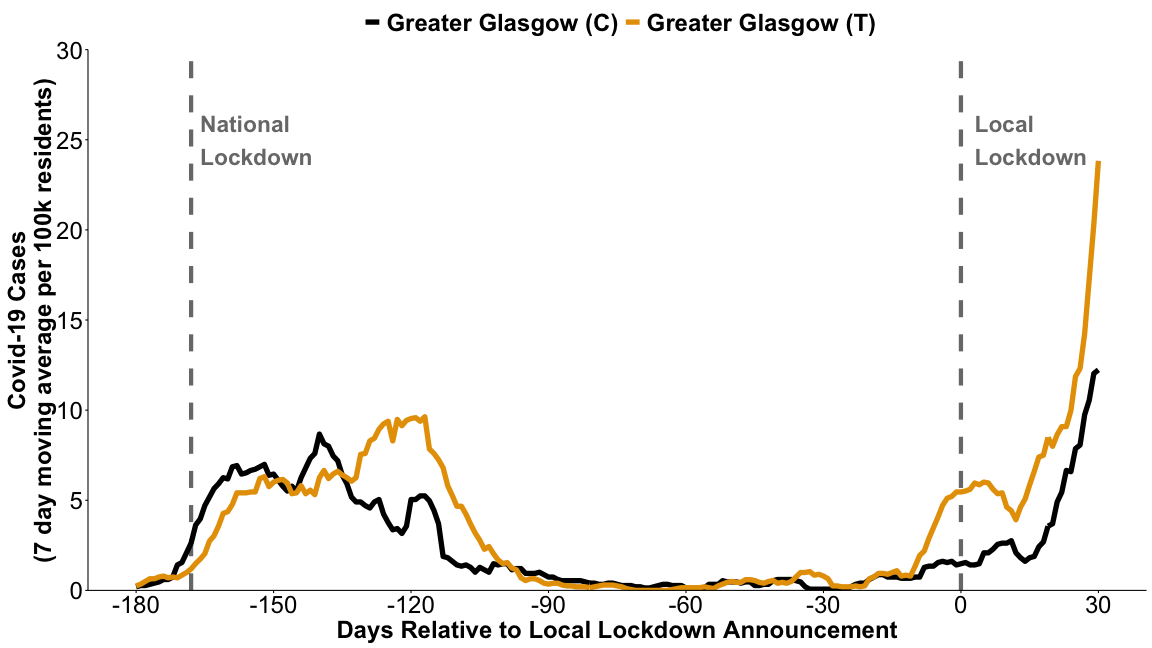}} \\ 
		\textbf{G. Lanarkshire} &
		\textbf{H. Leicester} &
		\textbf{I. Preston} \\
		{\includegraphics[height=1in]{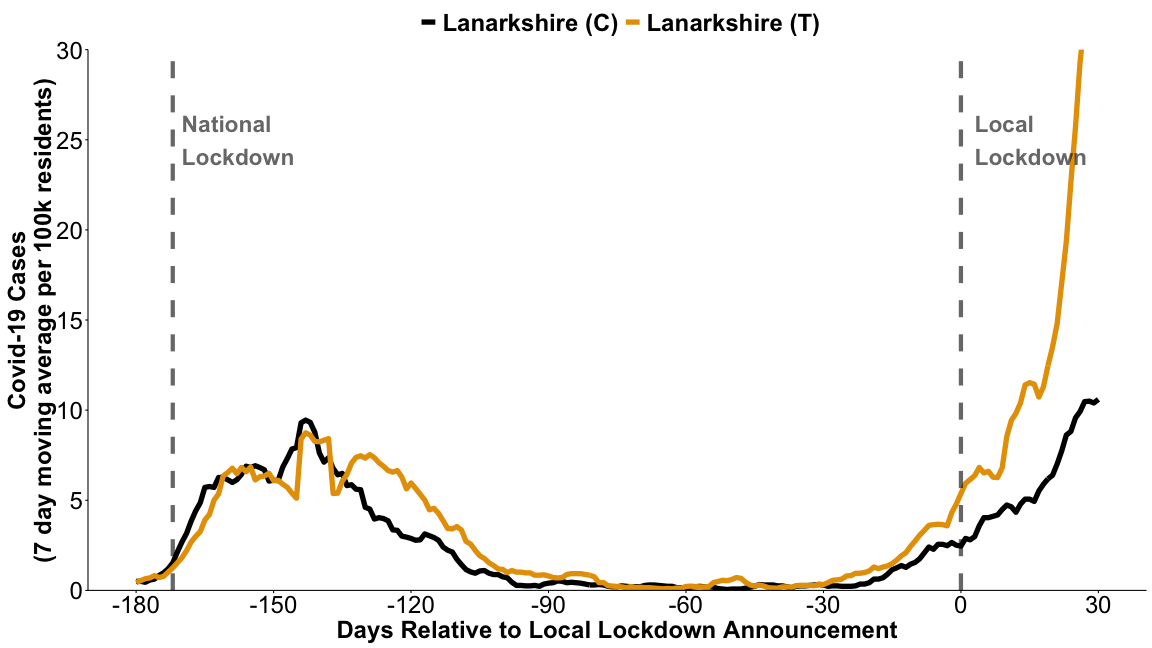}} & {\includegraphics[height=1in]{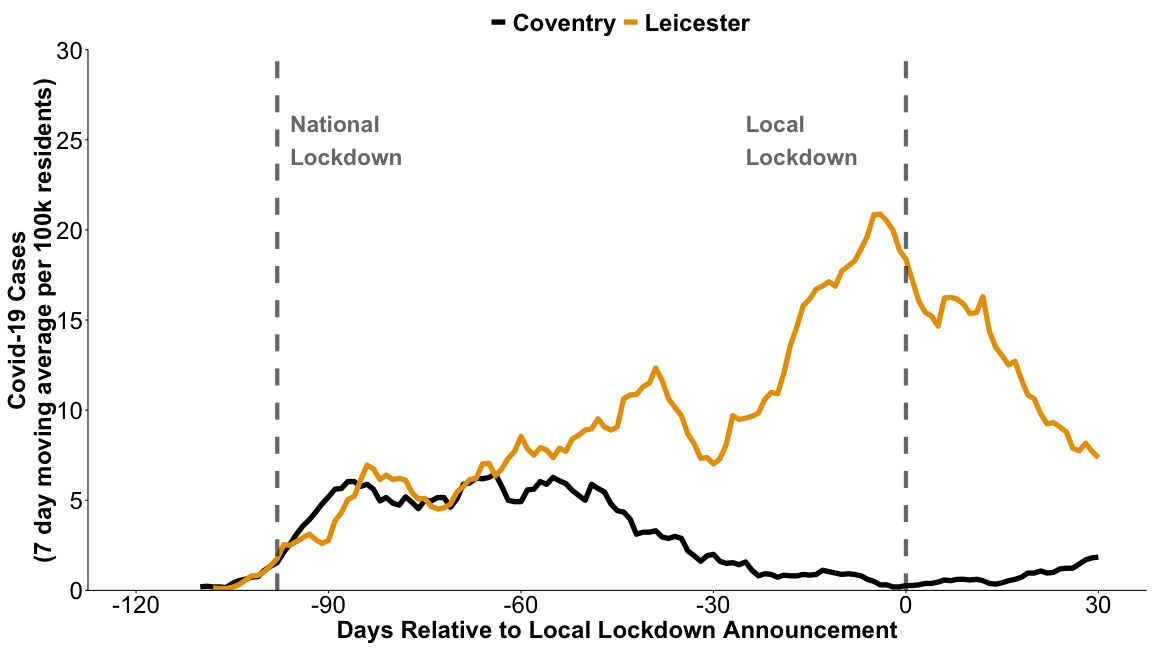}} & {\includegraphics[height=1in]{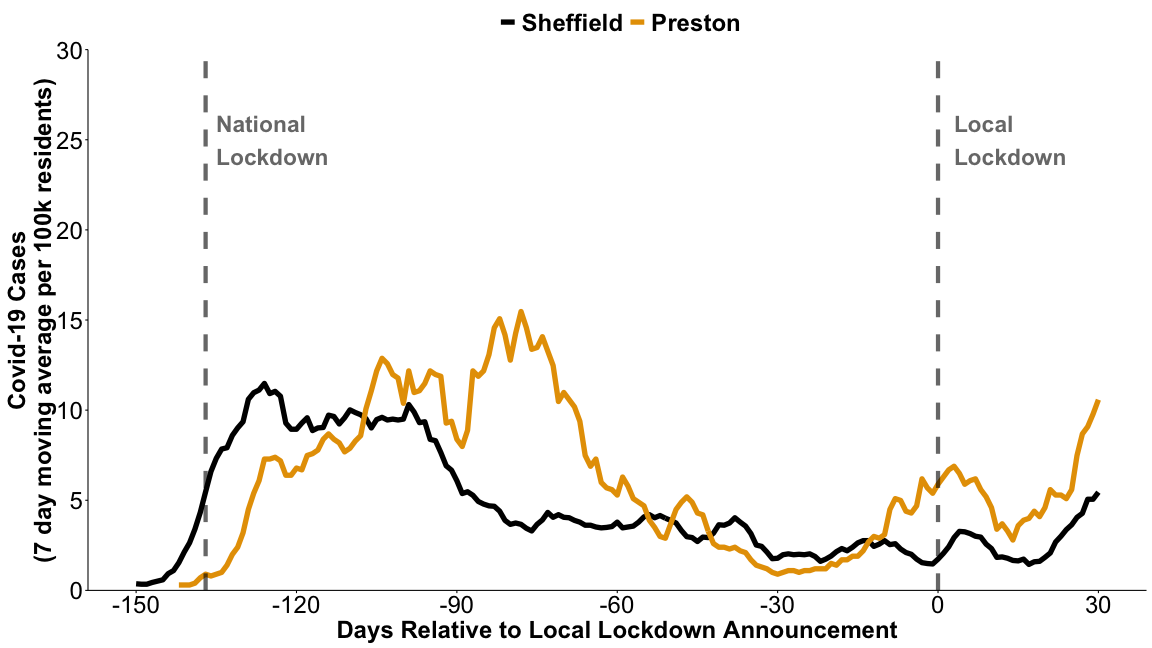}} \\ 
		\textbf{J. Leeds} &
		\textbf{K. Newcastle} &
		\textbf{L. Wolverhampton} \\
		{\includegraphics[height=1in]{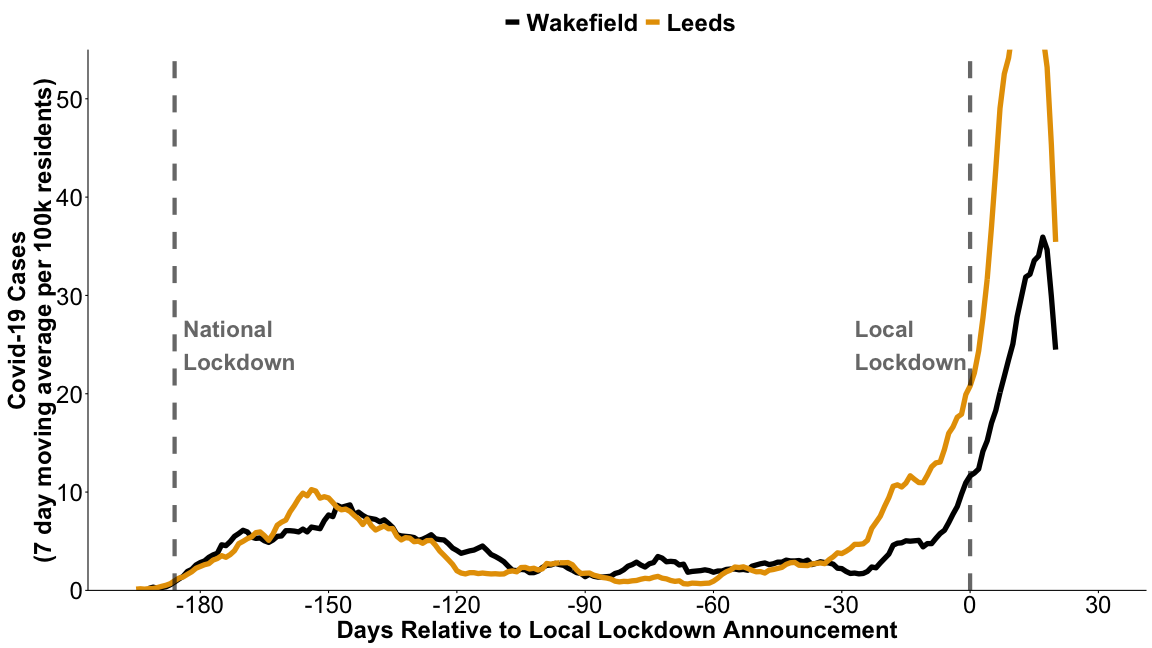}} & {\includegraphics[height=1in]{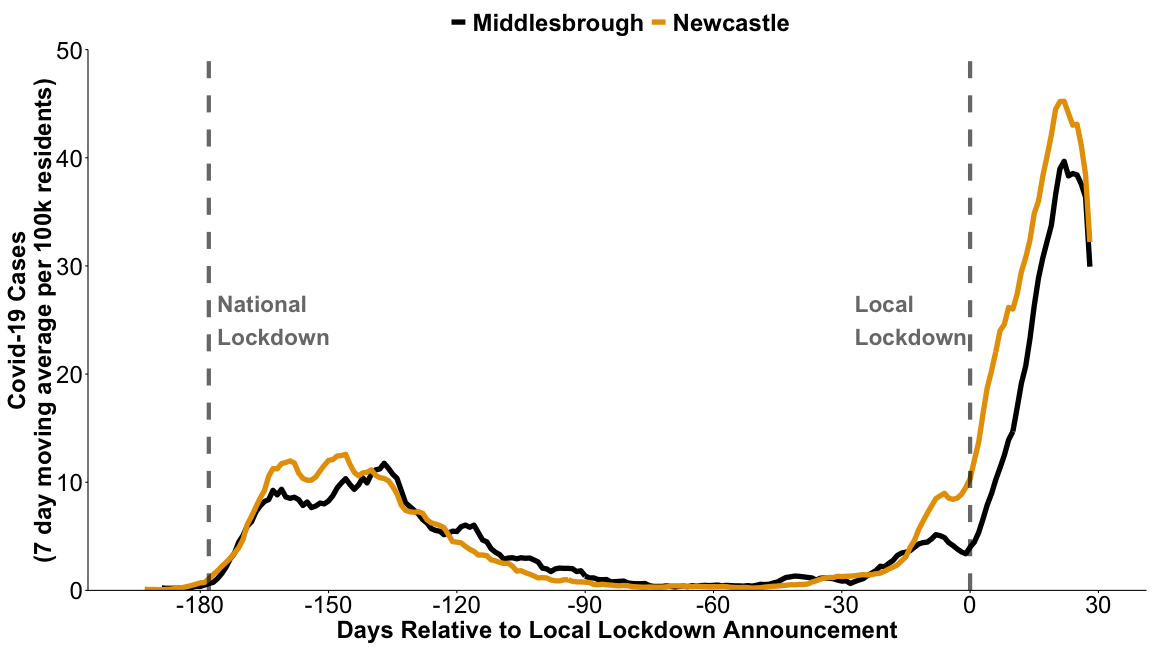}} & {\includegraphics[height=1in]{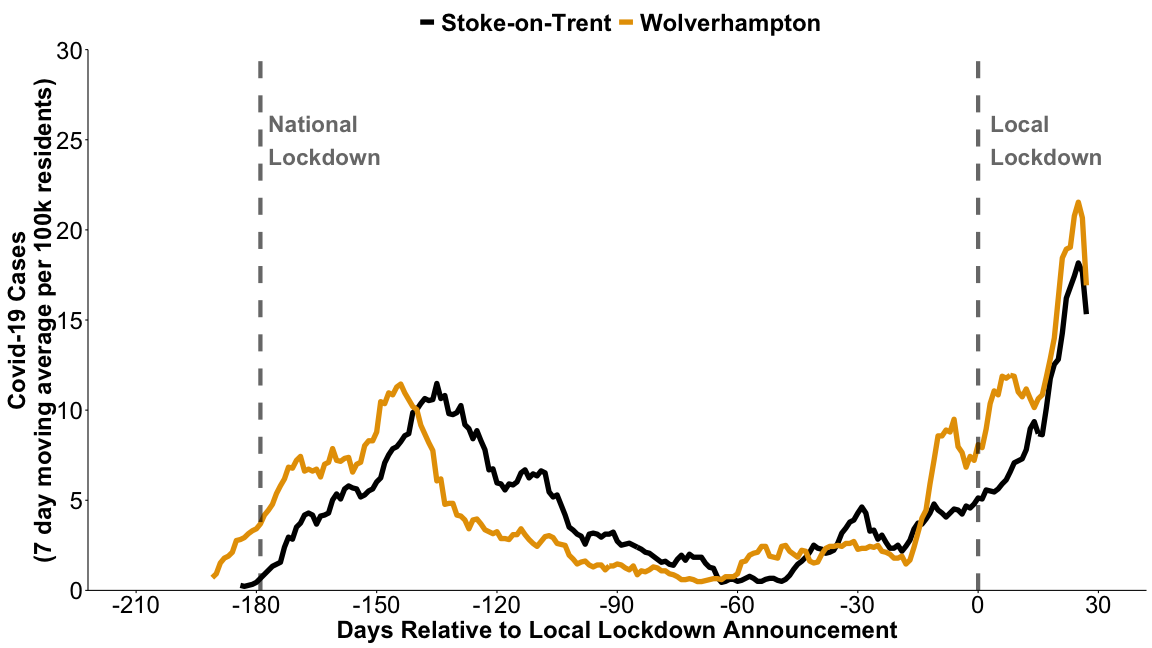}} \\ 
	\end{tabular}
	\label{fig:covidgrid}
	\begin{tablenotes}
		\small
		\item \textit{Notes: 7 day moving averages. Data from Public Health England, Public Health Wales and Public Health Scotland.}
	\end{tablenotes}
\end{figure}

We first examine the effects of local lockdowns on COVID-19 cases. The descriptive results are displayed in Figure \ref{fig:covidgrid}. Vertical dotted lines display the timing of the national lockdown in March 2020 (affecting both treatment and control groups) and the local lockdown (day 0 when it was announced) that only affects the (yellow) treatment areas not the (black) control localities.

There are clear pre-lockdown differences in case data between lockdown areas and comparison areas, with lockdowns being introduced following a sharp increase in the case rate in each of the city graphs shown.
The higher pre-lockdown case rate in lockdown areas may partially reflect a policy choice by health authorities whereby more tests are purposefully carried out in areas the government is considering introducing local lockdown restrictions. This is consistent with what we observe in COVID-19 positive cases as a percent of tests, however, such data are only available weekly and at a higher level of geographic region (upper rather than lower tier local authority) that does not align to areas subject to local lockdowns.

We observe cases continue to rise after the lockdown announcement -- as would be expected given the disease's incubation period -- but typically find that following the lockdown announcement the rise in cases ceases, case numbers peak and then level-off or decline.
This therefore indicates that local lockdowns can be effective at containing COVID-19 outbreaks.
However, while this is generally the case the results are heterogeneous with some exceptions to this – most notably Bolton and Glasgow where cases continued to rise after local lockdowns were imposed.

We estimate the relationship between local lockdowns and the evolution of cases using a difference-in-difference regression model. While our estimates should be interpreted as descriptive, we limit the sample for lockdown cases for which the pre-lockdown data indicates that the comparison geography appears to be a relatively suitable control for the treatment locality.\footnote{\cite{roth2018pre} highlights the limitations of relying on common pre-trends tests as evidence for common trends assumption for estimates to be interpreted as causal.}
This leads us to drop Bolton and Leicester as there were far sharper rises in COVID-19 incidences before the lockdown than for their control groups.
We also drop Aberdeen and Caerphilly as we have a relatively small number of transactions for these areas. Thus we have a remaining sample of six local lockdowns to study: Manchester, Birmingham, Glasgow, Greater Glasgow, Newcastle and Preston.

The dynamic regression results in Table 1 quantifies the relationship between local lockdowns and COVID-19 cases. In each case, the coefficient on the local lockdown dummy the first week post-lockdown is positive, and in most cases remains positive after two weeks, consistent with local lockdowns being imposed as cases rise towards a spike. The coefficients at subsequent time horizons fall, in the case of Manchester becoming negative in period one month after the imposition of the local lockdown.

\newpage

\newpage

\begin{table}[!htbp] \centering 
  \caption{Dynamic Difference-in-Difference Estimate on COVID-19 cases per 100,000 residents (7 day moving average)} 
  \label{} 
\begin{tabular}{@{\extracolsep{-10pt}}lcccccc} 
\\[-1.8ex]\hline 
\hline \\[-1.8ex] 
 & \multicolumn{6}{c}{\textit{Dependent variable:}} \\ 
\cline{2-7} 
\\[-1.8ex] & \multicolumn{6}{c}{COVID-19 cases per 100,000 residents} \\ 
 & Manchester & Preston & Glasgow & G. Glasgow & Birmingham & Newcastle \\ 
\\[-1.8ex] & (1) & (2) & (3) & (4) & (5) & (6)\\ 
\hline \\[-1.8ex] 
 $Treat \text{*} After\textsubscript{-3}$ & $-$2.015$^{***}$ & $-$4.104$^{***}$ & $-$1.171$^{***}$ & $-$2.874$^{***}$ & $-$4.129$^{***}$ & $-$4.381$^{***}$ \\ 
  & (0.141) & (0.434) & (0.254) & (0.350) & (0.656) & (0.399) \\ 
  & & & & & & \\ 
 $Treat \text{*} After\textsubscript{-2}$ & $-$1.475$^{***}$ & $-$3.866$^{***}$ & $-$0.927$^{***}$ & $-$2.868$^{***}$ & $-$4.836$^{***}$ & $-$5.223$^{***}$ \\ 
  & (0.117) & (0.429) & (0.245) & (0.348) & (0.636) & (0.373) \\ 
  & & & & & & \\ 
 $Treat \text{*} After\textsubscript{-1}$ & $-$0.850$^{***}$ & $-$2.566$^{***}$ & $-$0.866$^{**}$ & $-$2.434$^{***}$ & $-$3.795$^{***}$ & $-$2.714$^{***}$ \\ 
  & (0.156) & (0.646) & (0.247) & (0.381) & (0.640) & (0.629) \\ 
  & & & & & & \\ 
 $Treat \text{*} After\textsubscript{1}$ & 0.393$^{**}$ & 0.343 & 3.439$^{***}$ & 1.194$^{**}$ & 0.560 & 4.801$^{***}$ \\ 
  & (0.119) & (0.521) & (0.596) & (0.359) & (0.652) & (0.963) \\ 
  & & & & & & \\ 
 $Treat \text{*} After\textsubscript{2}$ & 1.102$^{***}$ & $-$1.054$^{*}$ & 4.251$^{***}$ & $-$0.470 & 4.867$^{***}$ & 6.700$^{***}$ \\ 
  & (0.163) & (0.507) & (0.262) & (0.419) & (1.179) & (0.583) \\ 
  & & & & & & \\ 
 $Treat \text{*} After\textsubscript{3}$ & 0.710$^{**}$ & $-$1.023$^{*}$ & 4.345$^{***}$ & 1.530$^{***}$ & $-$7.994 & 3.446$^{***}$ \\ 
  & (0.261) & (0.497) & (0.455) & (0.406) & (6.084) & (0.466) \\ 
  & & & & & & \\ 
 $Treat \text{*} After\textsubscript{4}$ & $-$0.496$^{***}$ & $-$0.442 & 11.720$^{***}$ & 0.777 & $-$87.443$^{***}$ &  \\ 
  & (0.134) & (0.604) & (1.092) & (0.431) & (13.297) &  \\ 
  & & & & & & \\ 
\hline \\[-1.8ex] 
\hline 
\hline \\[-1.8ex] 
\textit{Note:}  & \multicolumn{6}{l}{$^{*}$p$<$0.05; $^{**}$p$<$0.01; $^{***}$p$<$0.001} \\ 
 & \multicolumn{6}{l}{OLS regression as specified in Equation 2 with fixed effects for areas} \\ 
 & \multicolumn{6}{l}{and days. Areas are weighted by their 2019 resident population.} \\ 
 & \multicolumn{6}{l}{Standard errors clustered at area level. Each column is for different} \\ 
 & \multicolumn{6}{l}{area subject to local lockdown (with its nearby control area).} \\ 
 & \multicolumn{6}{l}{Daily data for 4 weeks pre and post local lockdown(3 weeks post } \\ 
 & \multicolumn{6}{l}{for Newcastle). Outcome is 7 day moving average of COVID cases per} \\ 
 & \multicolumn{6}{l}{100,000 residents in area from Public Health England and Scotland.} \\ 
 & \multicolumn{6}{l}{Omitted category is week (days -7 to -1) preceeding lockdown.} \\ 
\end{tabular} 
\end{table}

\subsection{Consumption}

\begin{figure}[H]
\caption{\textbf{Credit Card Spending in Manchester (yellow) and Liverpool (black)}} 
\centering
\begin{tabular}{c} 
\textbf{A. Overall} \\
{\includegraphics[height=2in]{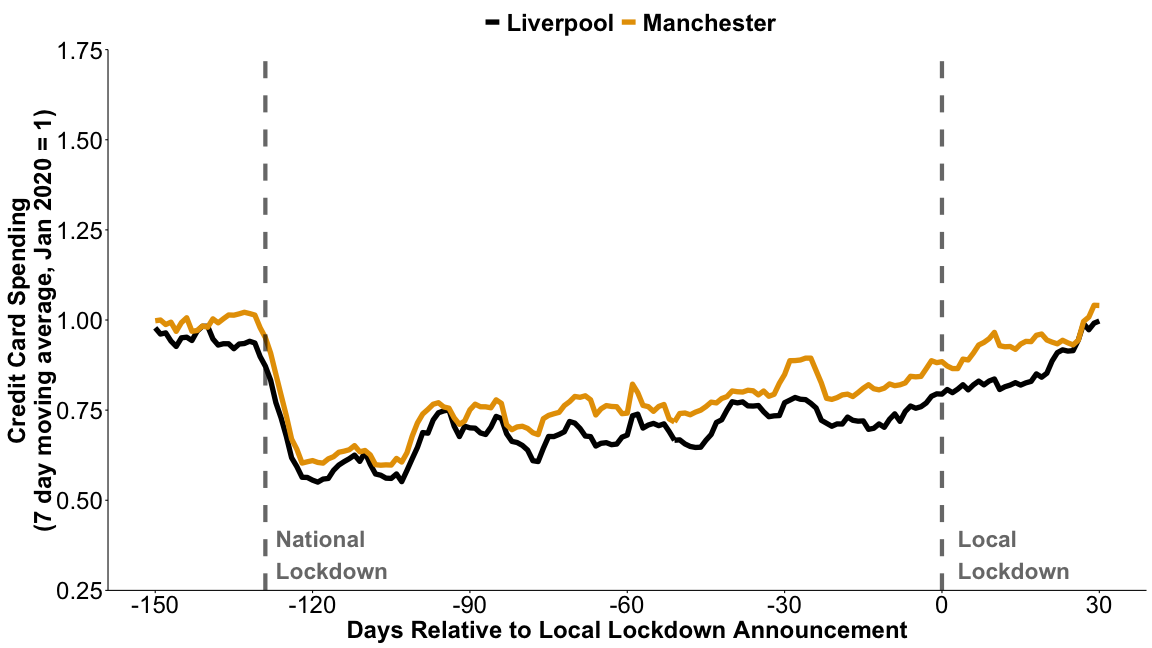}} \\ 
\textbf{B. Offline} \\
{\includegraphics[height=2in]{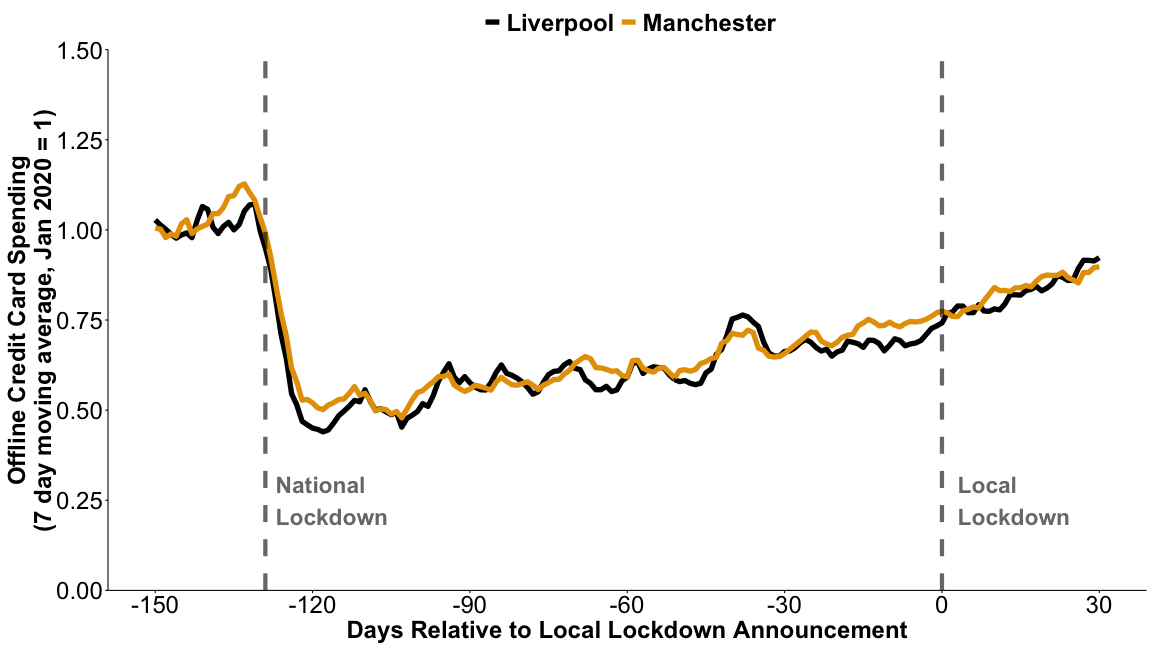}} \\ 
\textbf{C. Food and beverage} \\
{\includegraphics[height=2in]{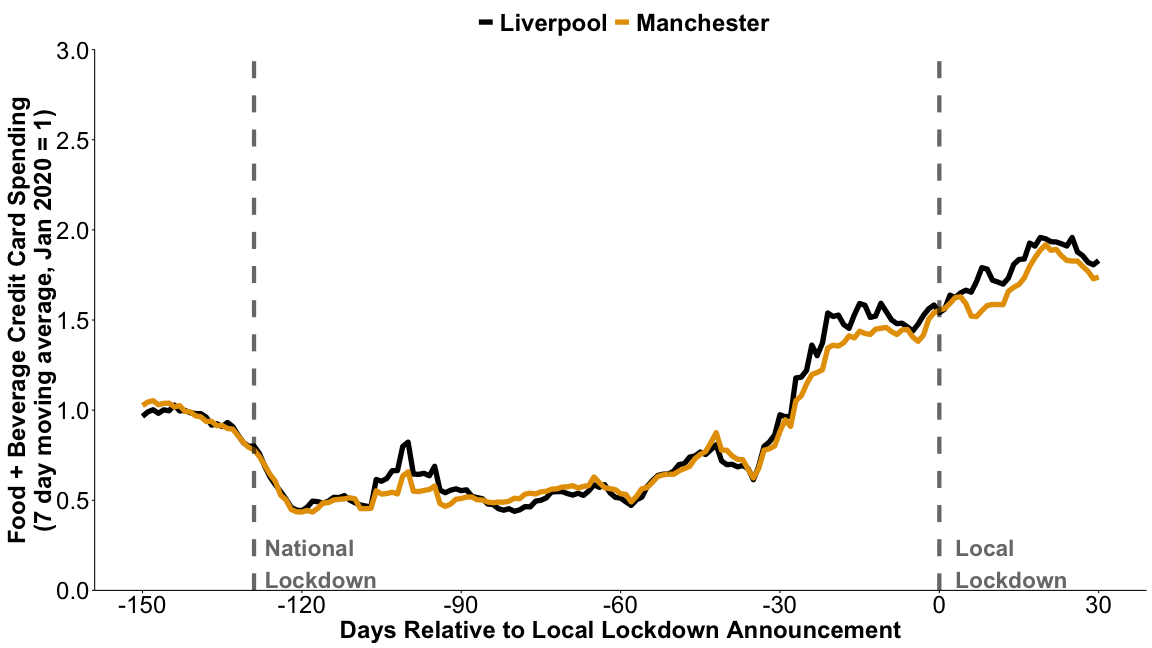}} \\ 
\textbf{D. Large store chains}\\ 
{\includegraphics[height=2in]{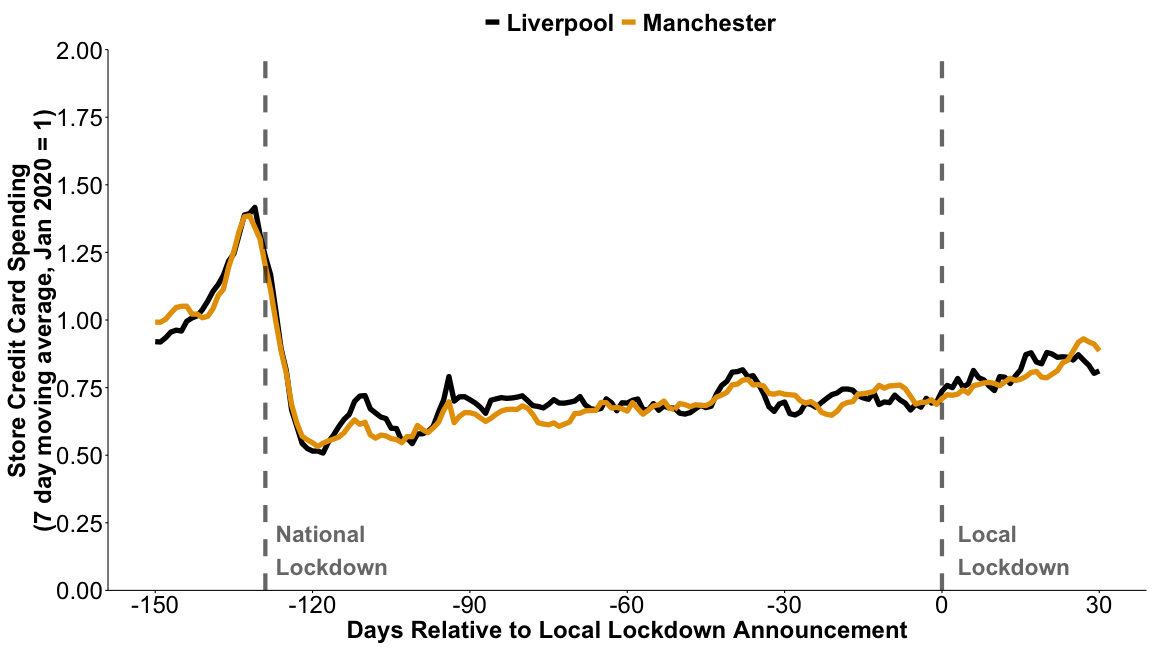}} \\ 
\end{tabular} \\ 
\label{fig:manchester}
\begin{tablenotes}
\small
\item \textit{Notes: Fable Data. Credit card spending measures use a 7 day moving average de-seasoned by taking ratio of the 7 day moving average a year prior. The series is then indexed to its moving average 8 - 28 January 2020. Panels A,B,C assign based on account-holder location. Panel D assigns based on large retail store chain location.} 
\end{tablenotes}
\end{figure}

\begin{figure}[H]
\caption{\textbf{Credit Card Spending in Lockdown Cities (yellow) vs Comparison Cities (black)}} 
\centering
\begin{tabular}{c c c} \\
\textbf{A. Aberdeen} &
\textbf{B. Birmingham} &
\textbf{C. Bolton} \\
{\includegraphics[height=1in]{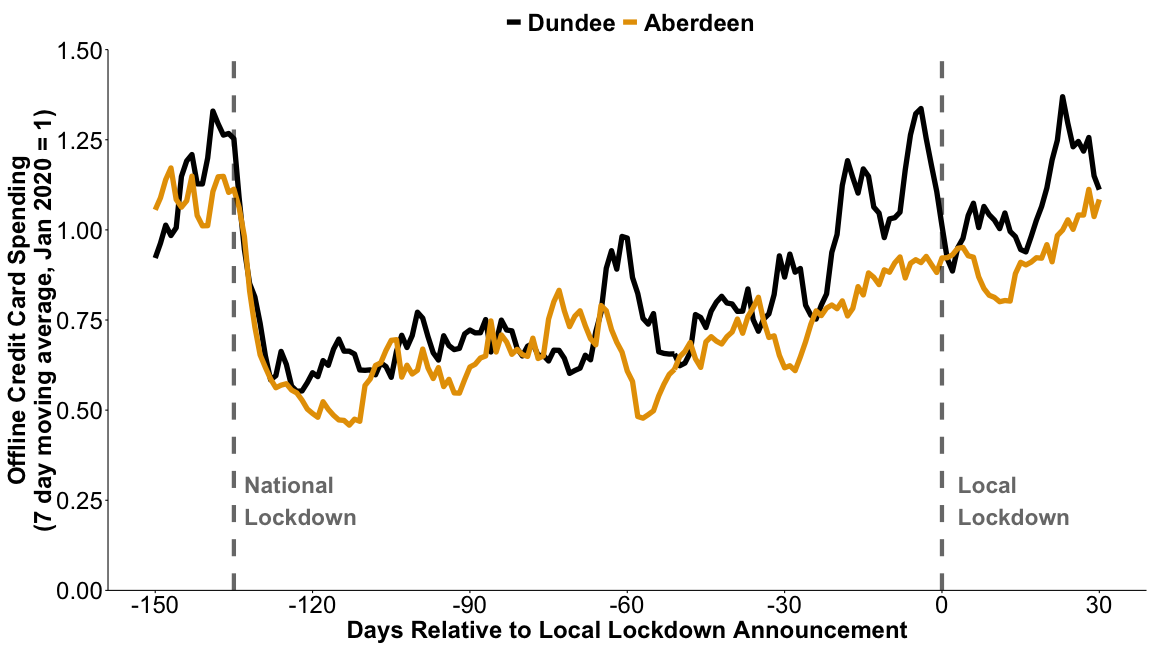}} & {\includegraphics[height=1in]{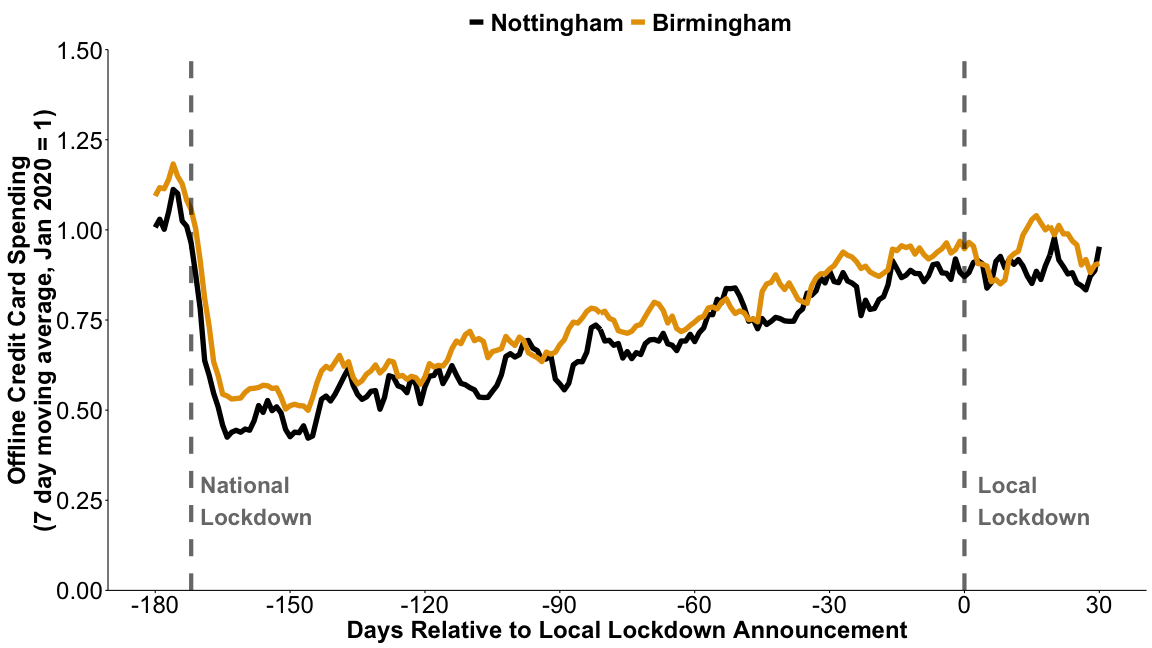}} & {\includegraphics[height=1in]{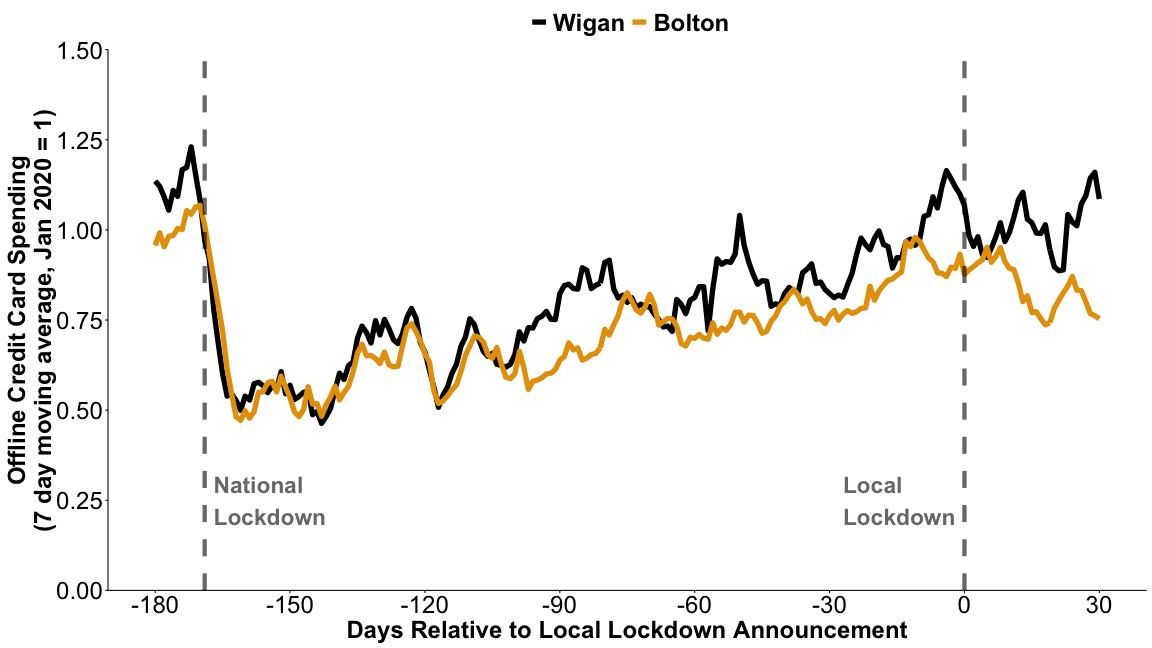}} \\ 
\textbf{D. Caerphilly} &
\textbf{E. Glasgow} &
\textbf{F. Greater Glasgow} \\
{\includegraphics[height=1in]{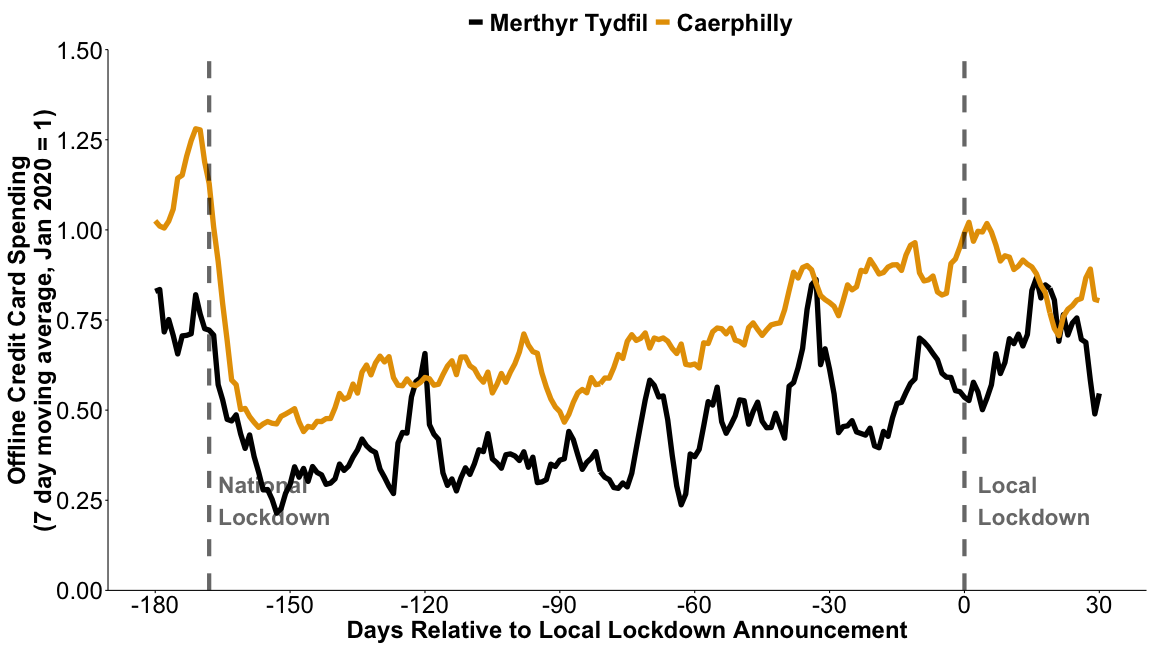}} & {\includegraphics[height=1in]{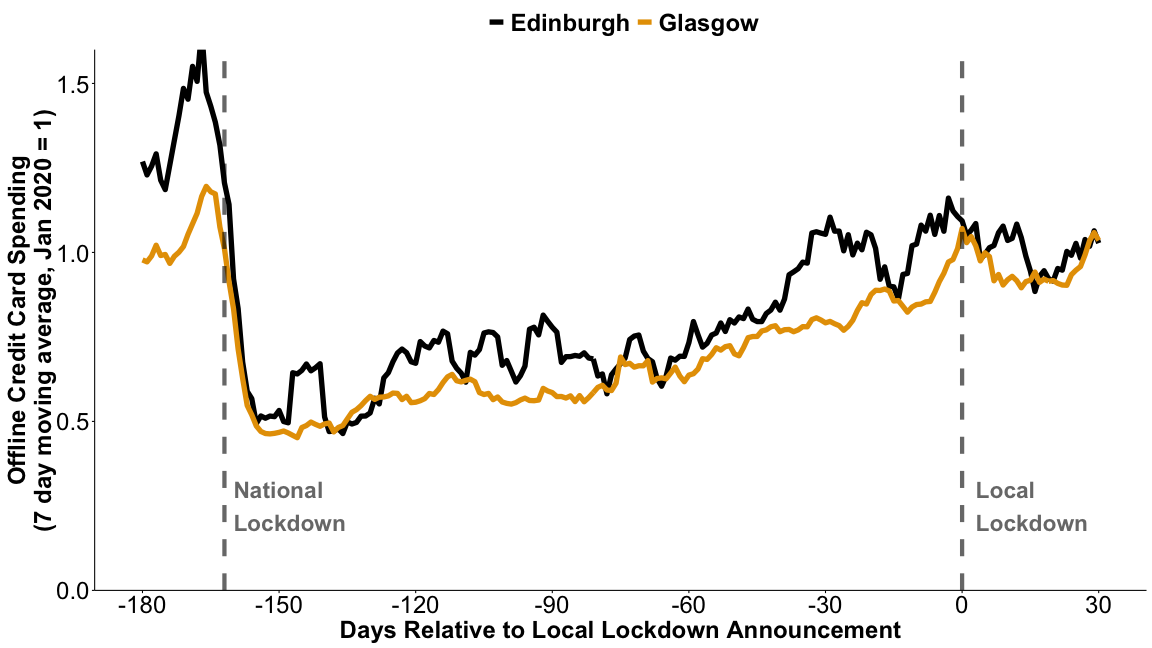}} & {\includegraphics[height=1in]{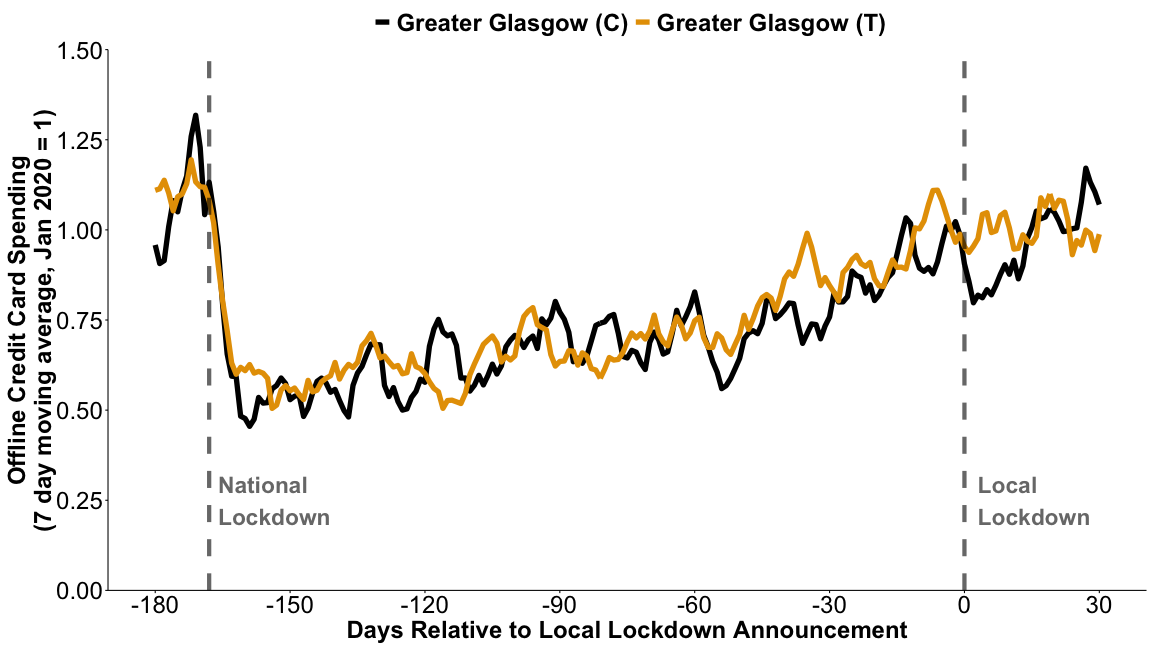}} \\ 
\textbf{G. Lanarkshire} &
\textbf{H. Leicester} &
\textbf{I. Preston} \\
{\includegraphics[height=1in]{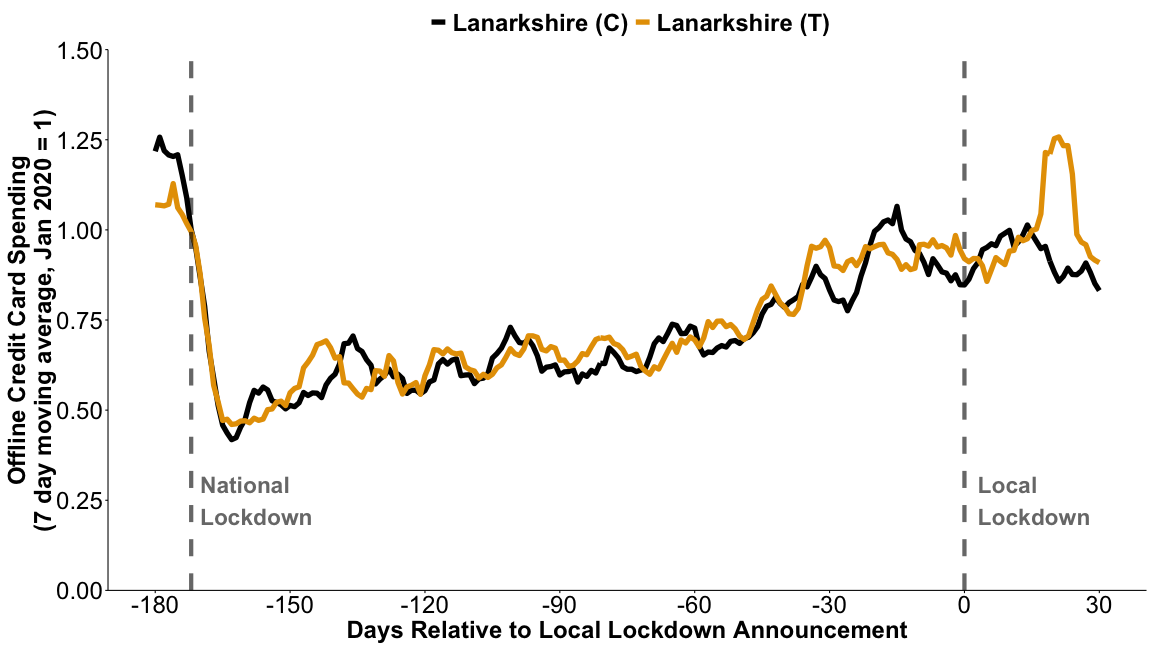}} & {\includegraphics[height=1in]{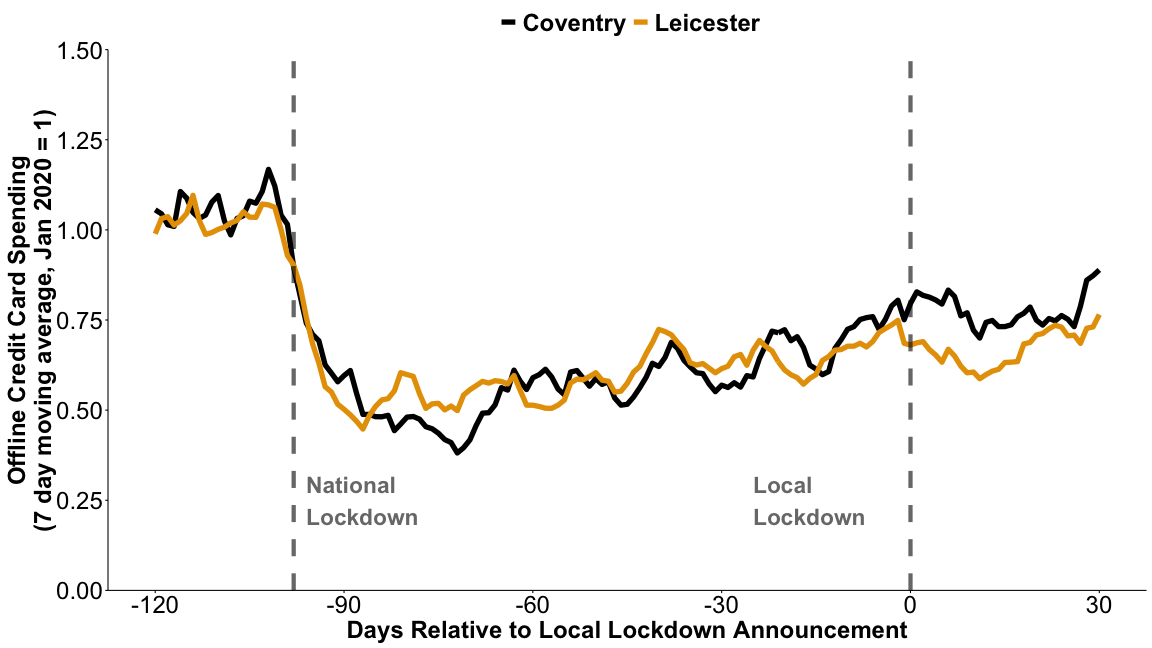}} & {\includegraphics[height=1in]{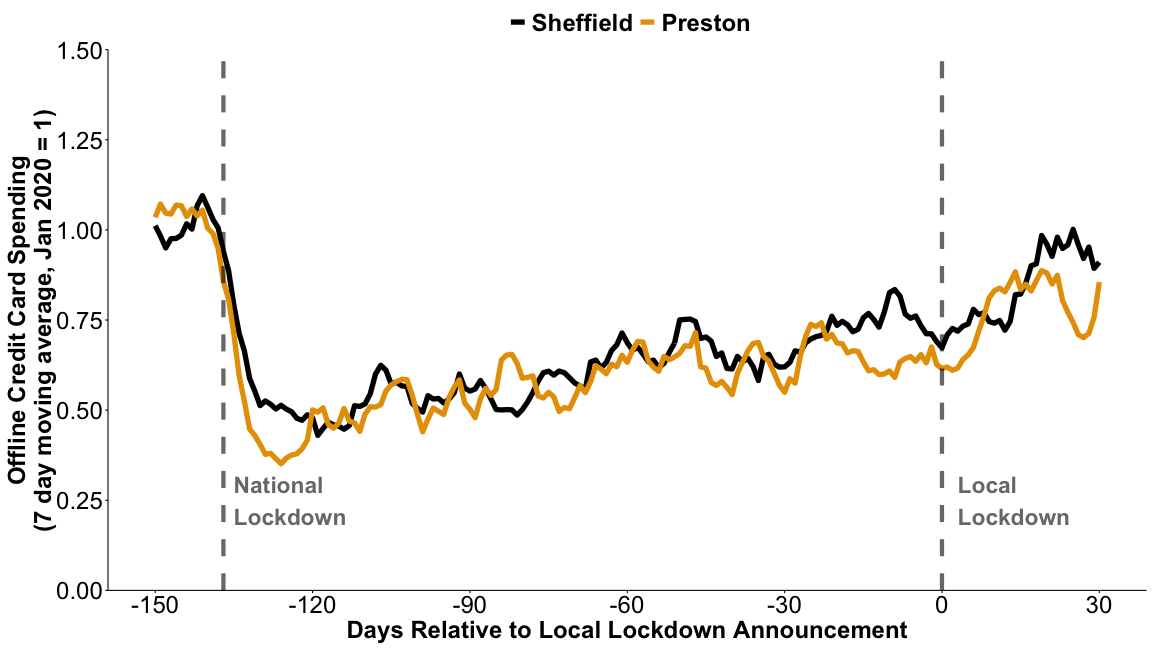}} \\ 
\textbf{J. Leeds} &
\textbf{K. Newcastle} &
\textbf{L. Wolverhampton} \\
{\includegraphics[height=1in]{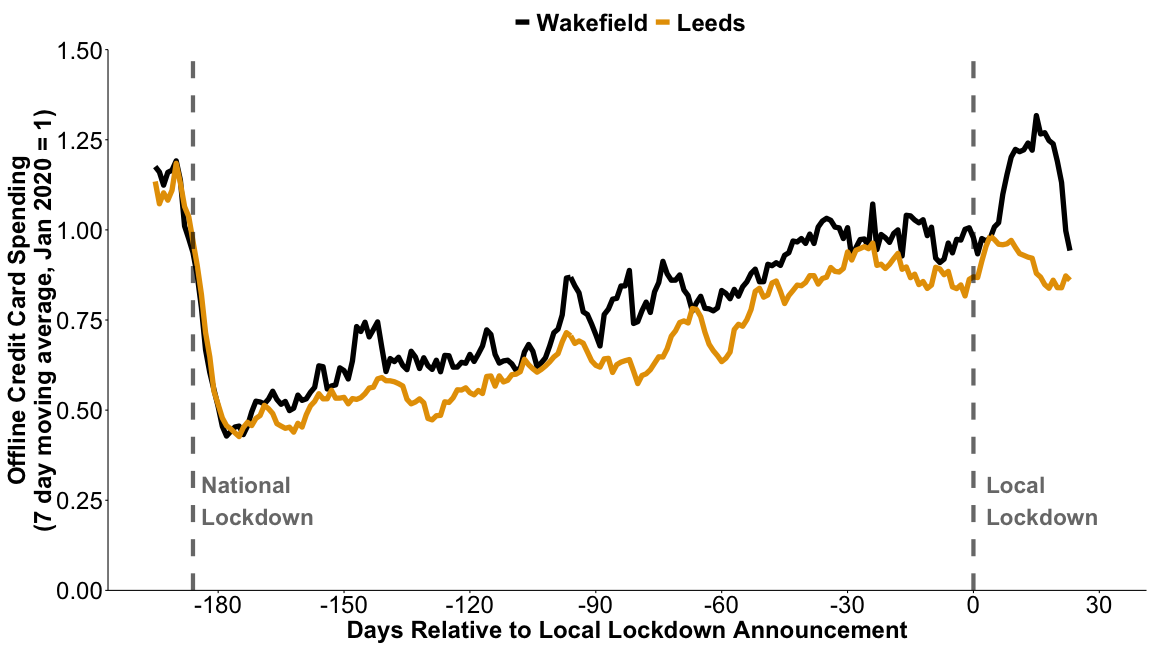}} & {\includegraphics[height=1in]{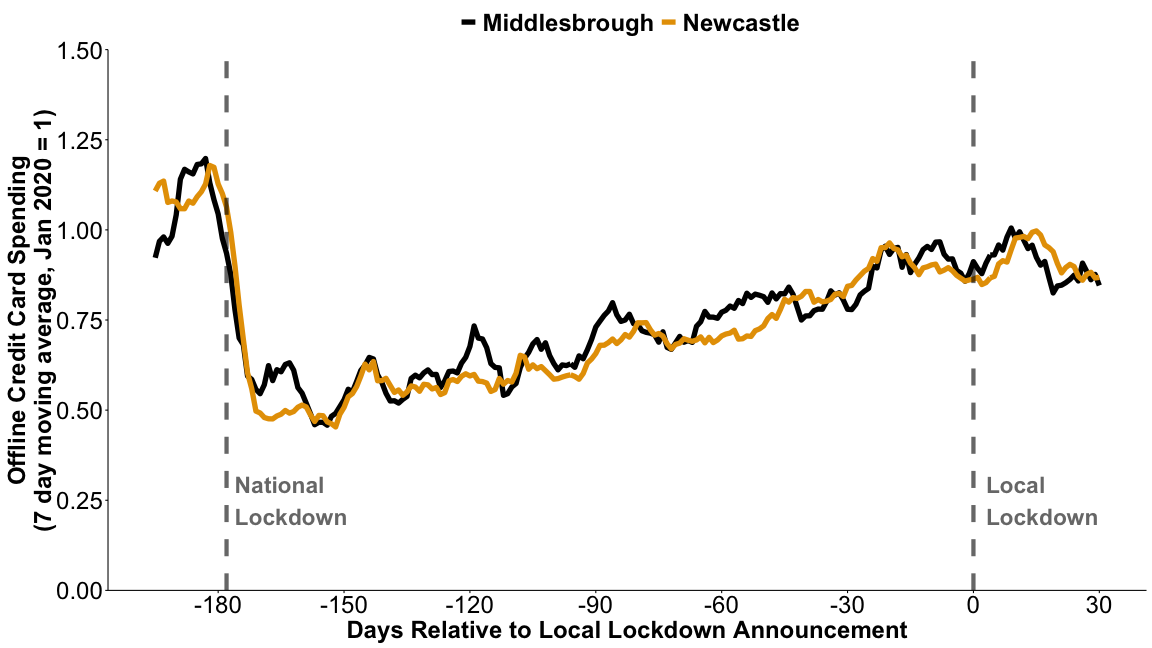}} & {\includegraphics[height=1in]{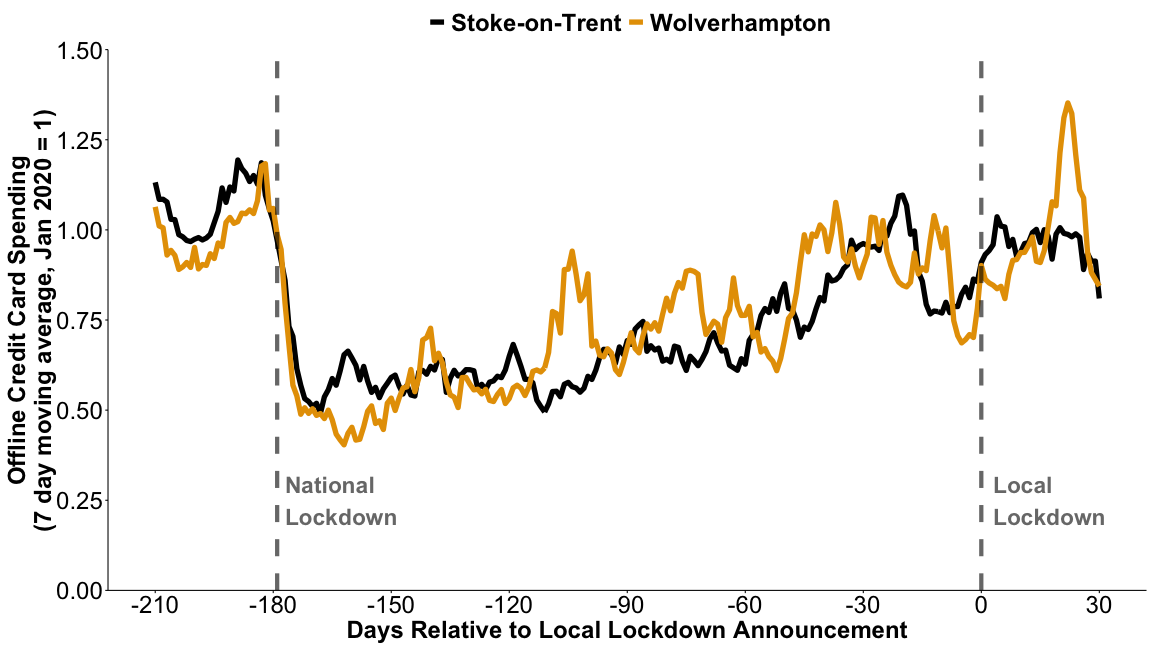}} \\
\end{tabular}
\label{fig:offgrid}
\begin{tablenotes}
\small
\item \textit{Notes: Fable Data. Offline credit card spending is a 7 day moving average de-seasoned by taking ratio of the 7 day moving average a year prior. The series is then indexed to its moving average 8 - 28 January 2020.}
\end{tablenotes}
\end{figure}

Figure \ref{fig:manchester} summarizes credit card spending for the Manchester local lockdown compared to the Liverpool control group. Each panel displays different spending measures (using 7 day moving averages): A. all credit card spending, B. offline credit card spending C. food and beverage credit card spending D. credit card spending in large store chains.
Our findings are consistent across these measures.
A broader set of treatment and control groups pairs are displayed for the same credit card spending measures in Figure \ref{fig:offgrid} (offline) and Annex Figures \ref{fig:allgrid} (all), \ref{fig:fbgrid} (food and beverage) and \ref{fig:storegrid} (large store chains, 14 day moving average) - due to smaller sample sizes in some of these areas the series are more volatile but show consistent results.

We highlight three features from these data.
First, the treatment and control regions have similarly-timed and sized declines in spending in March 2020.
The declines in consumption are economically large though we caveat that there is no control group unaffected by the virus or national lockdown to estimate the causal effects.
Second, the treatment and control regions typically have similar trends in consumption in the lead up to the local lockdown being announced.
These two features provide supportive evidence that our controls provide reasonable comparisons for the treatment areas.

Third, we observe little, if any, spending declines following the local lockdowns.
While there is heterogeneity in results with differences in signs and statistical significance across pairs, a consistent feature is that we can rule out there being any economically large declines of the magnitude clearly observed (nationally or for these localities) in March 2020.

This descriptive analysis therefore indicates that local lockdowns are not having the large negative effect on consumption that the first wave of the virus and national lockdown did.
Our dynamic regression results displayed in Table 2 also show this, however, we caveat it as in some weeks, in some lockdowns spending is lower. For example in Manchester, it is no more so than the difference observed pre-lockdown and certainly nowhere near the spending declines accompanying the March 2020 national shutdown.

\newpage

\newpage 

\newpage

\begin{table}[!htbp] \centering 
  \caption{Dynamic Difference-in-Difference Estimate on Offline credit card spending (7 day moving average)} 
  \label{} 
\begin{tabular}{@{\extracolsep{-5pt}}lcccccc} 
\\[-1.8ex]\hline 
\hline \\[-1.8ex] 
 & \multicolumn{6}{c}{\textit{Dependent variable:}} \\ 
\cline{2-7} 
\\[-1.8ex] & \multicolumn{6}{c}{Offline credit card spending} \\ 
 & Manchester & Preston & Glasgow & G. Glasgow & Birmingham & Newcastle \\ 
\\[-1.8ex] & (1) & (2) & (3) & (4) & (5) & (6)\\ 
\hline \\[-1.8ex] 
 $Treat \text{*} After\textsubscript{-3}$ & $-$0.031$^{***}$ & 0.088$^{***}$ & $-$0.062 & $-$0.023 & 0.024 & 0.061$^{***}$ \\ 
  & (0.008) & (0.025) & (0.032) & (0.043) & (0.017) & (0.012) \\ 
  & & & & & & \\ 
 $Treat \text{*} After\textsubscript{-2}$ & $-$0.015 & 0.020 & 0.072 & $-$0.052 & 0.009 & 0.032$^{**}$ \\ 
  & (0.009) & (0.018) & (0.042) & (0.044) & (0.017) & (0.010) \\ 
  & & & & & & \\ 
 $Treat \text{*} After\textsubscript{-1}$ & 0.003 & $-$0.090$^{***}$ & 0.023 & $-$0.048 & 0.012 & $-$0.029$^{*}$ \\ 
  & (0.007) & (0.023) & (0.043) & (0.064) & (0.014) & (0.011) \\ 
  & & & & & & \\ 
 $Treat \text{*} After\textsubscript{1}$ & $-$0.049$^{***}$ & $-$0.005 & 0.145$^{***}$ & 0.076 & $-$0.019 & $-$0.027$^{**}$ \\ 
  & (0.010) & (0.017) & (0.026) & (0.049) & (0.022) & (0.009) \\ 
  & & & & & & \\ 
 $Treat \text{*} After\textsubscript{2}$ & $-$0.018 & 0.146$^{***}$ & 0.032 & 0.038 & $-$0.059$^{*}$ & 0.001 \\ 
  & (0.011) & (0.031) & (0.027) & (0.046) & (0.028) & (0.017) \\ 
  & & & & & & \\ 
 $Treat \text{*} After\textsubscript{3}$ & $-$0.028$^{***}$ & 0.056 & 0.154$^{***}$ & $-$0.073 & 0.060$^{*}$ & 0.092$^{***}$ \\ 
  & (0.007) & (0.029) & (0.029) & (0.045) & (0.029) & (0.013) \\ 
  & & & & & & \\ 
 $Treat \text{*} After\textsubscript{4}$ & $-$0.054$^{***}$ & $-$0.090$^{*}$ & 0.111$^{***}$ & $-$0.107 & 0.034$^{*}$ &  \\ 
  & (0.011) & (0.034) & (0.026) & (0.055) & (0.015) &  \\ 
  & & & & & & \\ 
\hline \\[-1.8ex] 
\hline 
\hline \\[-1.8ex] 
\textit{Note:}  & \multicolumn{6}{l}{$^{*}$p$<$0.05; $^{**}$p$<$0.01; $^{***}$p$<$0.001} \\ 
 & \multicolumn{6}{l}{OLS regression as specified in Equation 2 with fixed effects for areas} \\ 
 & \multicolumn{6}{l}{and days. Areas are weighted by their 2019 resident population.} \\ 
 & \multicolumn{6}{l}{Standard errors clustered at area level. Each column is } \\ 
 & \multicolumn{6}{l}{for different area subject to local lockdown (with its nearby} \\ 
 & \multicolumn{6}{l}{control area). Daily data for 4 weeks pre and post local lockdown} \\ 
 & \multicolumn{6}{l}{(3 weeks post for Newcastle). Outcome is Fable Data daily } \\ 
 & \multicolumn{6}{l}{series index for 7 day moving average of offline consumer credit} \\ 
 & \multicolumn{6}{l}{card spending. Outcome is de-seasoned by taking ratio of 7 day} \\ 
 & \multicolumn{6}{l}{moving average a year prior and indexed (=1)to its moving} \\ 
 & \multicolumn{6}{l}{average 8-28 January 2020. Omitted category is week (days -7} \\ 
 & \multicolumn{6}{l}{to -1) preceeding lockdown.} \\ 
\end{tabular} 
\end{table}

\newpage

\section{Conclusions}

We introduce a new real-time source of consumption data that we demonstrate is a highly correlated, leading indicator of official statistics, is also available at transaction level and can be disaggregated to produce daily measures across geographies.

Our analysis studies how consumer spending responds to UK local lockdowns using a difference-in-difference approach, comparing nearby cities or small areas.
We do not find large spending declines in response to these local lockdowns.
Instead we find little (if any) decline in local spending.
To help interpret these estimates we observe that they are far smaller than the credit card spending declines observed nationally or for these localities during the combination of the first virus wave and national lockdown in March 2020.
Using the same difference-in-difference methodology we find that these local lockdowns typically appear to turn the tide on rising COVID-19 positive cases.

We thus conclude that it appears possible for policymakers to use such local lockdowns restricting household mixing to contain COVID-19 outbreaks without killing local economies.
However, we caveat this by noting that the effectiveness of such measures to mitigate the virus itself are expected to be highly dependent on an effective system of testing to isolate which regions to lockdown and contain infected individuals.
It is also expected to depend upon co-ordination across regions by governments to ensure outbreaks in one area are contained and do not spillover into other areas.

While our evidence indicates some initial successes from local lockdowns, the UK (along with many European countries) experienced a rapid, nationwide rise in COVID-19 cases in late September and October following the end of the summer holidays, reopening of schools and universities (though the cause of the wave's sudden rise is not yet clear) indicating that its system of nationwide containment is not isolating cases early enough to be effective.

In Autumn and Winter 2020 a series of rapid COVID-19 outbreaks across the UK (and also other European countries) have since resulted in government imposing new, more restrictive systems of local and national lockdowns that includes forced business closures.
In \cite{gathergood2020lev}, we document early evidence on how the measures implemented to November 2020 had unequal effects across UK regions.
Whether measures being imposed in 2021 result in changes to consumption more like the first national lockdown or the local lockdowns we study here will have profoundly different implications for the UK economy's prospects and other countries facing virus outbreaks.

\singlespacing 
\bibliography{refs}
\bibliographystyle{apalike}
\doublespacing

\newpage

\section*{Annex}

\setcounter{figure}{0}
\renewcommand{\thefigure}{A\arabic{figure}}

\setcounter{table}{0}
\renewcommand{\thetable}{A\arabic{table}}

\begin{table}[H]
\centering
\caption{\textbf{Local lockdown announcement dates, areas \& controls}} 
\resizebox{\textwidth}{!}{%
\begin{tabular}{|c|c|c|} \hline
\textbf{Announcement Date} & \textbf{Local Lockdown} & \textbf{Control} \\ \hline
\textbf{29/6} & \textbf{Leicester} & \textbf{Coventry} \\ 
 & Leicester, Oadby and Wigston & Coventry \\ \hline
\textbf{30/7} & \textbf{Manchester} & \textbf{Liverpool} \\ 
 & Manchester, Bury, Oldham, Rochdale, Salford, & Liverpool, Halton, Knowsley, \\ 
 & Salford, Stockport, Tameside, Trafford, & St. Helens, Sefton, Wirral \\ 
 & Blackburn with Darwen, Bradford, Calderdale, & \\
 & Rossendale, Pendle, Hyndburn, Burnley & \\ \hline
\textbf{5/8} & \textbf{Aberdeen} & \textbf{Dundee} \\ 
 & Aberdeen City & Dundee City \\ \hline
\textbf{7/8} & \textbf{Preston} & \textbf{Sheffield} \\  
 & Preston & Sheffield \\ \hline 
\textbf{1/9} & \textbf{Glasgow} & \textbf{Edinburgh}  \\ 
 & Glasgow City, East Renfrewshire, & Edinburgh City, West Lothian  \\ 
 & West Dunbartonshire &  \\ \hline
\textbf{7/9} & \textbf{Greater Glasgow (T)} & \textbf{Greater Glasgow (C)} \\ 
 & Renfrewshire, East Dunbartonshire & Iverclyde, North Ayrshire \\ \hline
\textbf{7/9} & \textbf{Caerphilly} & \textbf{Merthyr Tydfil} \\  
 & Caerphilly & Merthyr Tydfil \\ \hline 
\textbf{5/9} & \textbf{Bolton} & \textbf{Wigan} \\  
 & Bolton & Wigan \\ \hline 
\textbf{11/9} & \textbf{Birmingham} & \textbf{Nottingham} \\  
 & Birmingham, Sandwell, Solihull & Nottingham \\ \hline 
\textbf{11/9} & \textbf{Lanarkshire (T)} & \textbf{Lanarkshire (C)} \\ 
 & South Lanarkshire, North Lanarkshire & Stirling, Falkirk, Scottish Borders, \\ 
 & & Midlothian, East Lothian \\ \hline
\textbf{11/9} & \textbf{Belfast} & \textbf{} \\ 
 & Belfast &  \\ \hline
\textbf{16/9} & \textbf{Rhondda Cynon Taf} & \textbf{} \\ 
 & Rhondda Cynon Taf &  \\ \hline
\textbf{17/9} & \textbf{Newcastle} & \textbf{Middlesbrough} \\ 
 & Newcastle upon Tyne, Gateshead, Sunderland, Northumberland, & Middlesbrough, Redcar and Cleveland, \\ 
 & South Tyneside, North Tyneside, County Durham & Stockton-on-Tees, Darlington \\ \hline
\textbf{18/9} & \textbf{Liverpool} & \textbf{} \\ 
 & Liverpool, Halton, Knowsley, &  \\
 & St. Helens, Sefton, Wirral &  \\ \hline
\textbf{18/9} & \textbf{Wolverhampton} & \textbf{Stoke-on-Trent} \\ 
 & Wolverhampton & Stoke-on-Trent \\ \hline
\textbf{18/9} & \textbf{Lancashire} & \textbf{} \\ 
 & Chorley, Flyde, Lancaster, Ribble Valley &  \\ 
 & South Ribble, West Lancashire, Wyre &  \\ \hline
\textbf{18/9} & \textbf{Warrington} & \textbf{} \\ 
 & Warrington &  \\ \hline
\textbf{18/9} & \textbf{Oadby and Wigston} & \textbf{} \\ 
 & Oadby and Wigston &  \\ \hline
\textbf{21/9} & \textbf{South Wales (1)} & \textbf{} \\ 
 & Blaenau Gwent, Bridgend, Merthyr Tydfil, Newport &  \\ \hline
\textbf{21/9} & \textbf{Northern Ireland} & \textbf{} \\ 
 & Rest of Northern Ireland &  \\
 & (Belfast already in lockdown) &  \\ \hline
\textbf{25/9} & \textbf{Blackpool} & \textbf{} \\ 
 & Blackpool &  \\ \hline
 \textbf{25/9} & \textbf{Leeds} & \textbf{Wakefield} \\ 
 & Leeds & Wakefield \\ \hline
\textbf{25/9} & \textbf{Stockport} & \textbf{} \\ 
 & Stockport &  \\ \hline
\textbf{25/9} & \textbf{Wigan} & \textbf{} \\ 
 & Wigan &  \\ \hline
\textbf{25/9} & \textbf{Welsh Cities} & \textbf{} \\ 
 & Cardiff, Swansea &  \\ \hline
\textbf{27/9} & \textbf{South Wales (2)} & \textbf{} \\ 
 & Neath Port Talbot, Torfaen, Vale of Glamorgan &  \\ \hline
\textbf{29/9} & \textbf{North Wales} & \textbf{} \\ 
 & Conwy, Denbighshire, Flintshire, Wrexham &  \\ \hline
\end{tabular}}
\begin{tablenotes}
\small
\item \textit{Notes: Lower tier local authorities listed.
This does not include areas below the local authority level (e.g. Blaby,  Charnwood, Carmarthenshire) where parts were locked down. `Control' lists lower tier local authorities chosen as control groups: blank where not used for analysis because region is small and/or no suitable control area exists. Bolton announced and immediately introduced requirements on 5/9 but a full local lockdown was subsequently announced on 8/9.}
\end{tablenotes}
\label{fig:dates}
\end{table}

\newpage

\begin{figure}[H]
	\centering
	\caption{\textbf{UK Credit Card Spending 2018 - 2020  }} 
	\vspace{1cm}
	\begin{tabular}{c} 
		\textbf{A. Fable \& Bank of England Monthly Data, 2018 - 2020} \\
		{\includegraphics[height=3in]{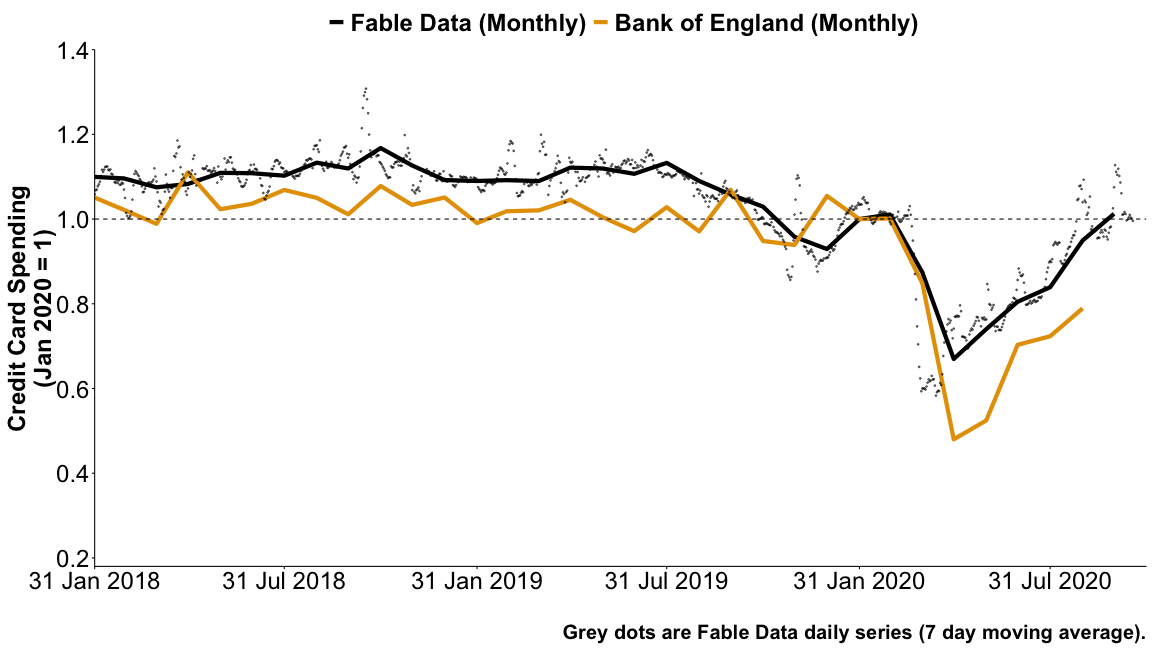}} \\ \\ 
		\textbf{B. Fable Daily Data, 2020} \\ \textbf{(7, 14, 28 day moving averages and monthly)} \\
		{\includegraphics[height=3in]{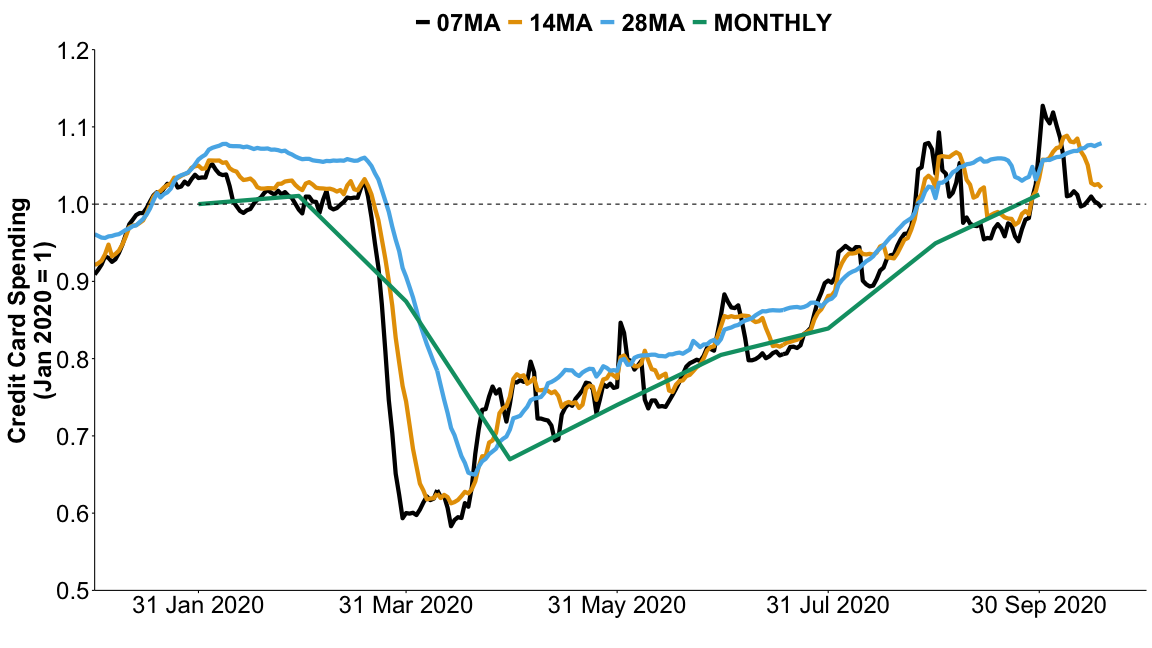}} 
	\end{tabular}
	\begin{tablenotes}
		\small
		\item \textit{Notes: Bank of England monthly data is derived from LPMVZQH (monthly gross credit card lending to individuals).
			Fable Data monthly series is indexed to January 2020.
			Fable Data 7,14,28 day moving averages are the daily moving average de-seasoned by taking ratio of the moving average a year prior. Each daily series is then indexed to its moving average 8 - 28 January 2020.
		}
	\end{tablenotes}
	\label{fig:boe}
\end{figure}

\newpage

\begin{figure}[H]
\caption{\textbf{Overall credit card spending in areas subject to local lockdown (yellow) compared to control areas not locked down (black), 7 day moving average}} 
\centering
\begin{tabular}{c c c} \\
\textbf{A. Aberdeen} &
\textbf{B. Birmingham} &
\textbf{C. Bolton} \\
{\includegraphics[height=1in]{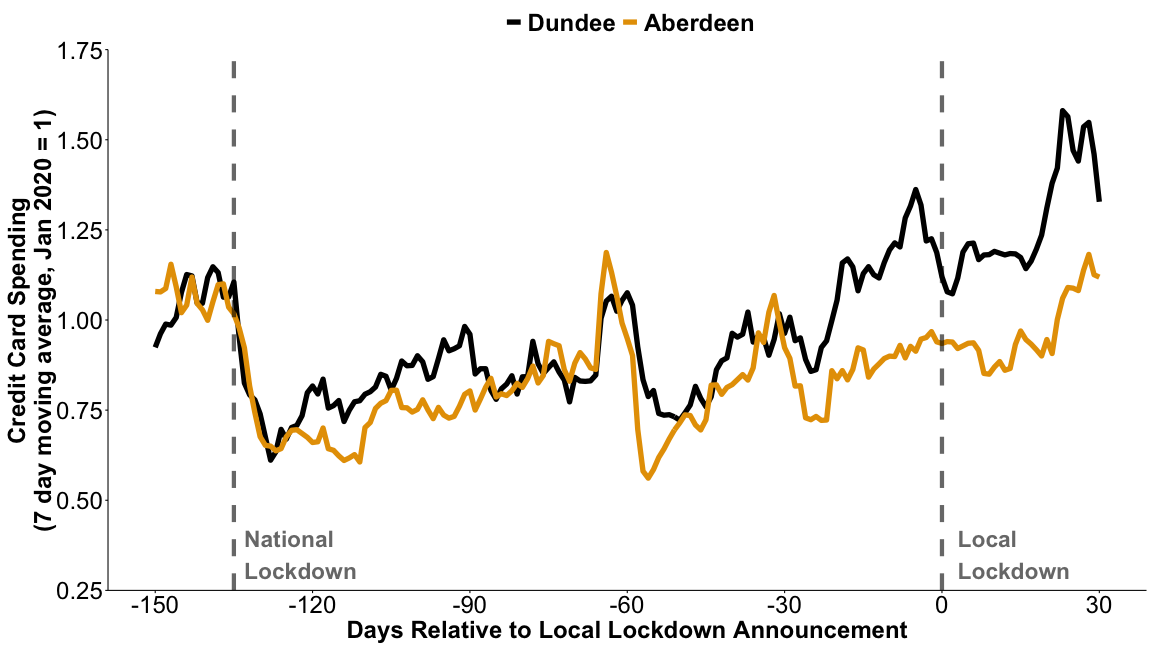}} & {\includegraphics[height=1in]{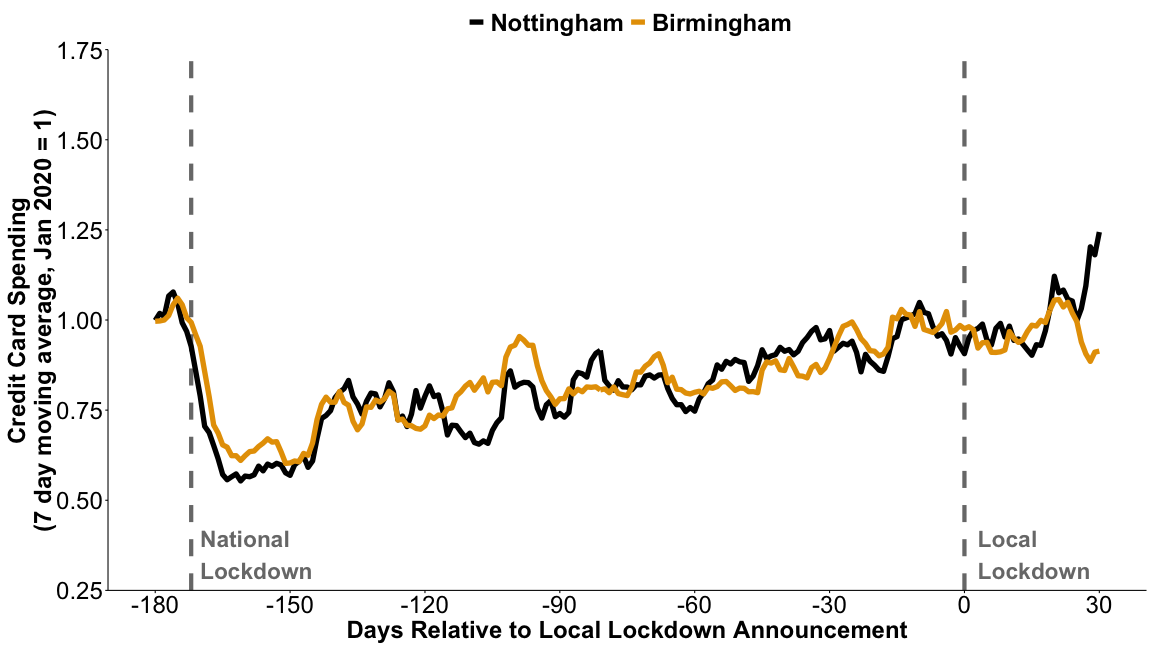}} & {\includegraphics[height=1in]{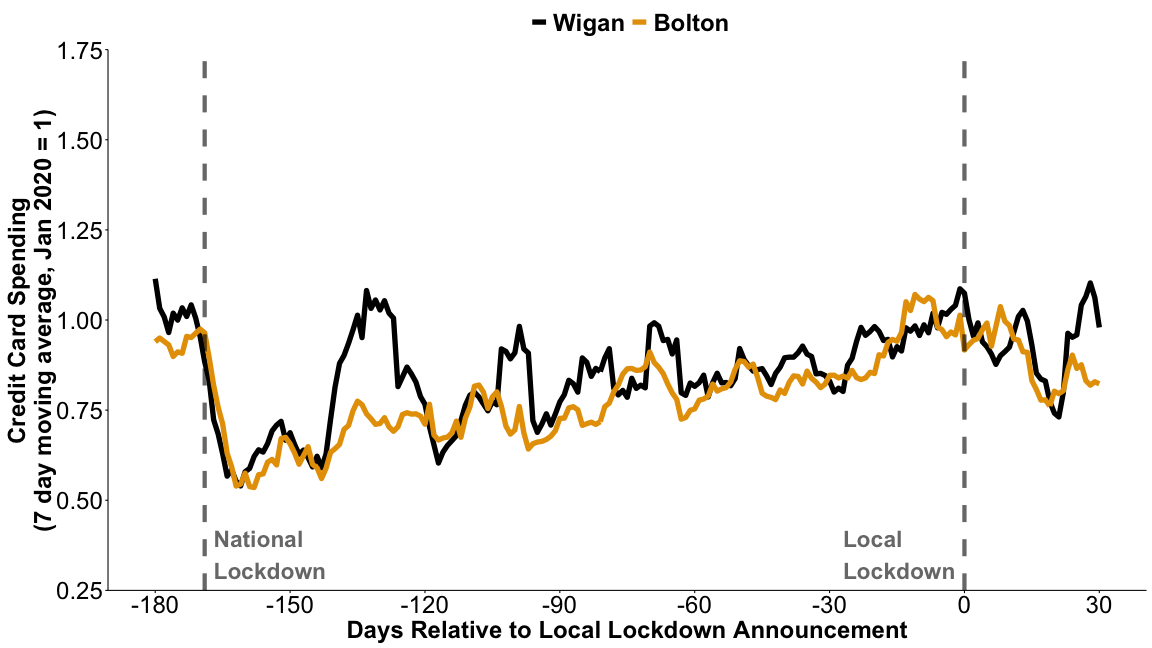}} \\ 
\textbf{D. Caerphilly} &
\textbf{E. Glasgow} &
\textbf{F. Greater Glasgow} \\
{\includegraphics[height=1in]{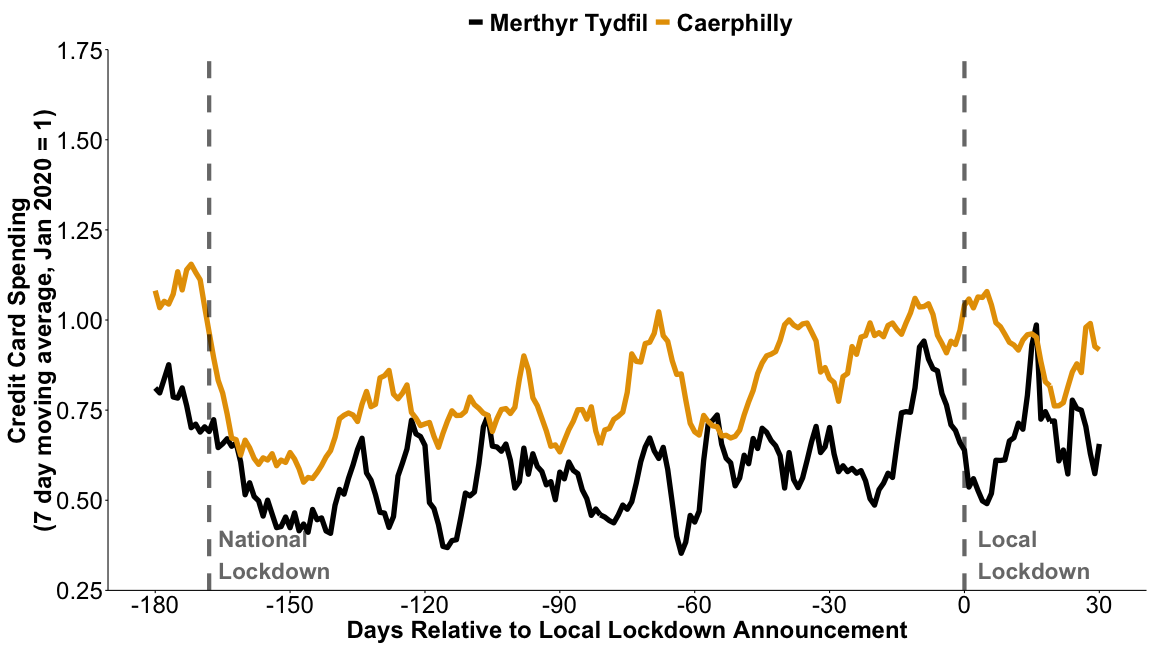}} & {\includegraphics[height=1in]{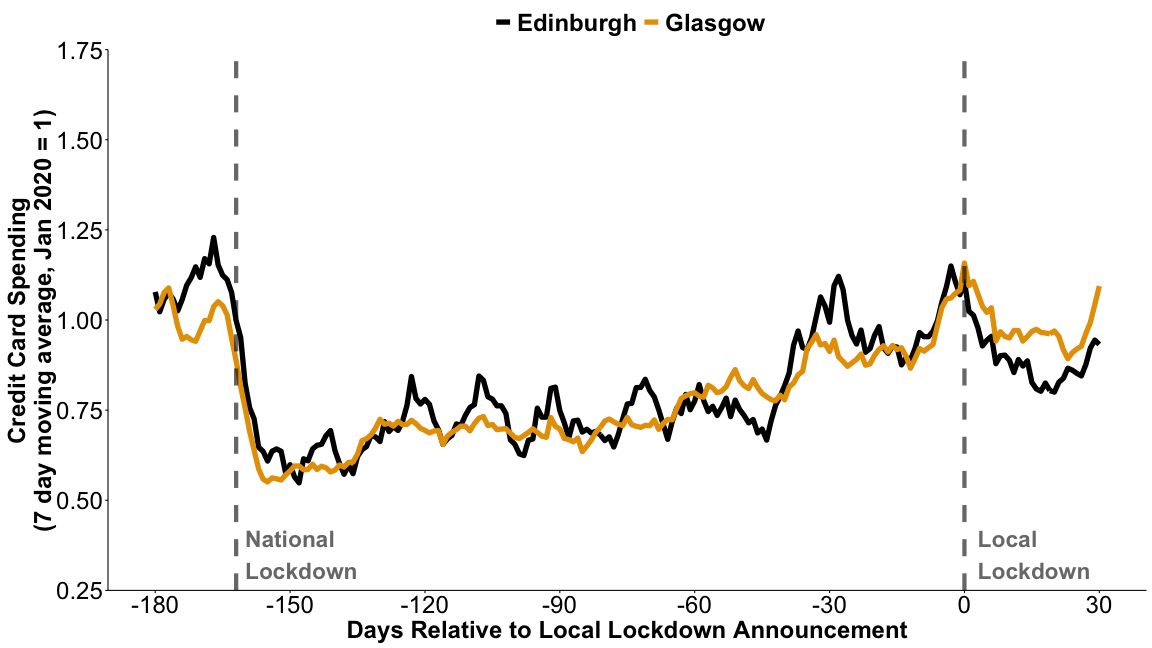}} & {\includegraphics[height=1in]{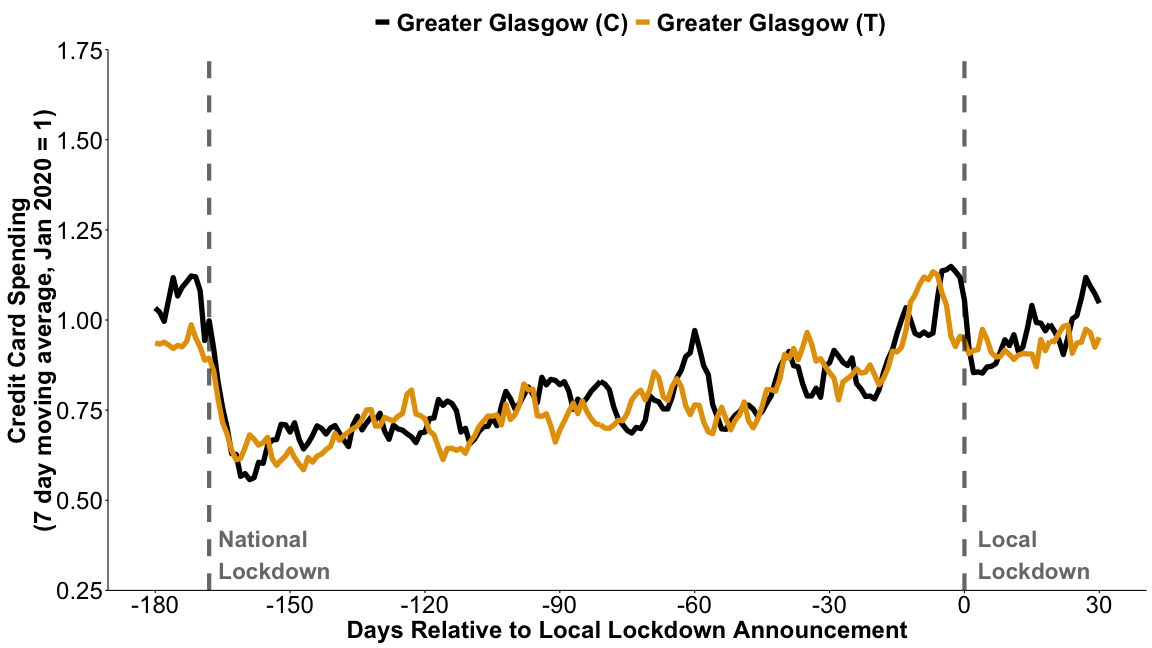}} \\ 
\textbf{G. Lanarkshire} &
\textbf{H. Leicester} &
\textbf{I. Preston} \\
{\includegraphics[height=1in]{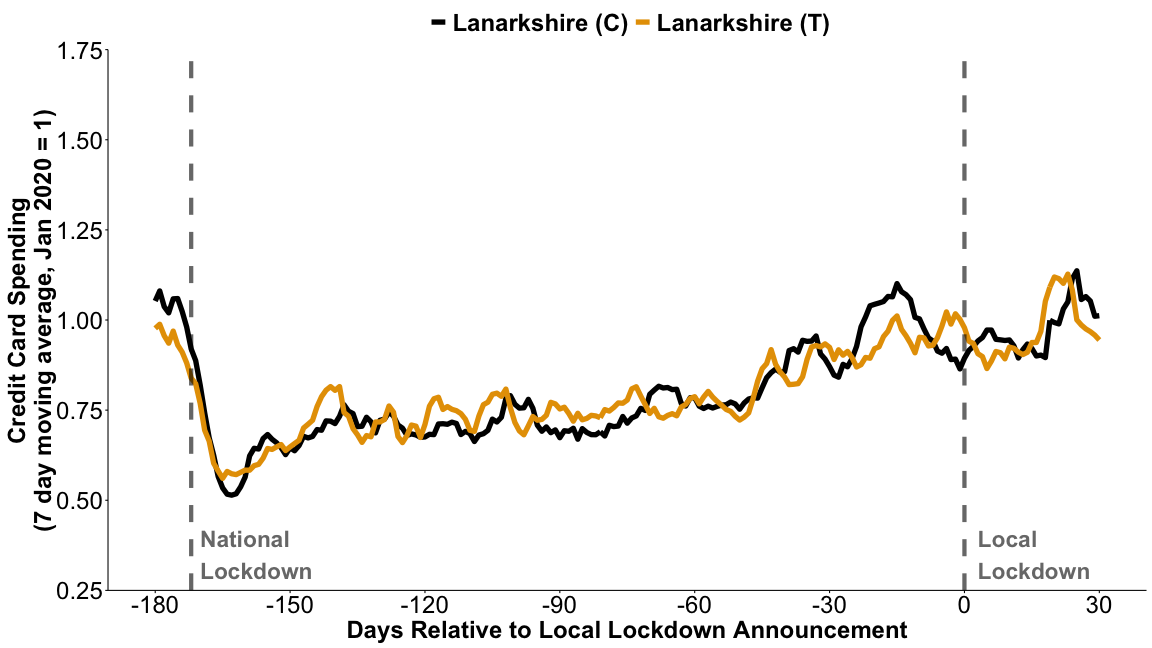}} & {\includegraphics[height=1in]{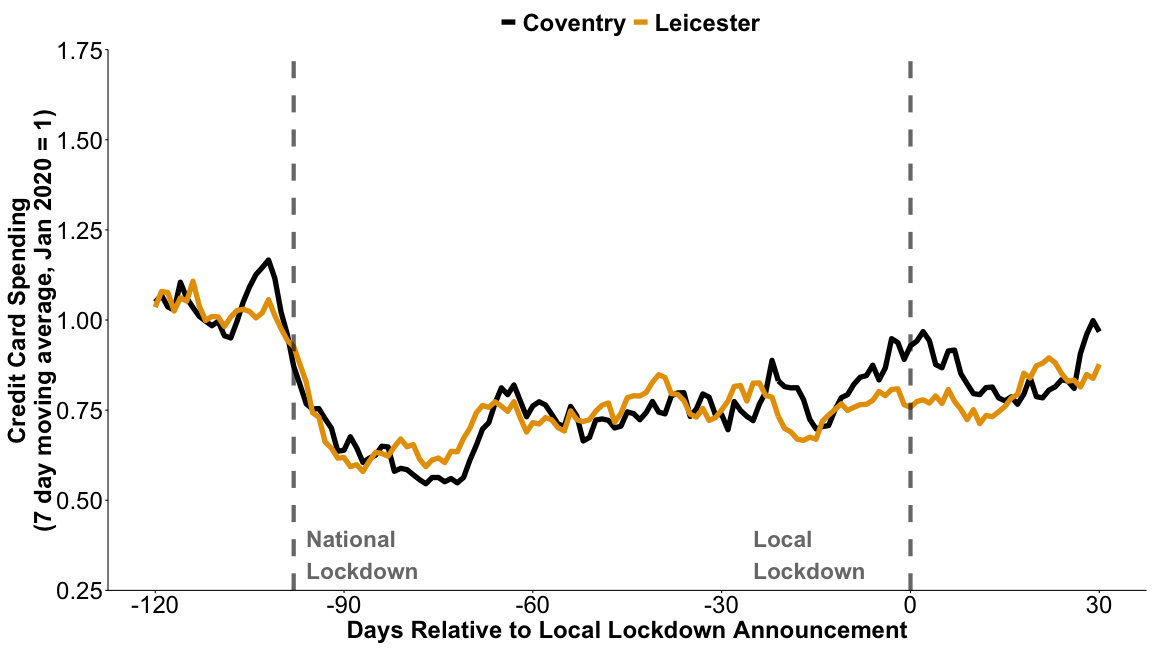}} & {\includegraphics[height=1in]{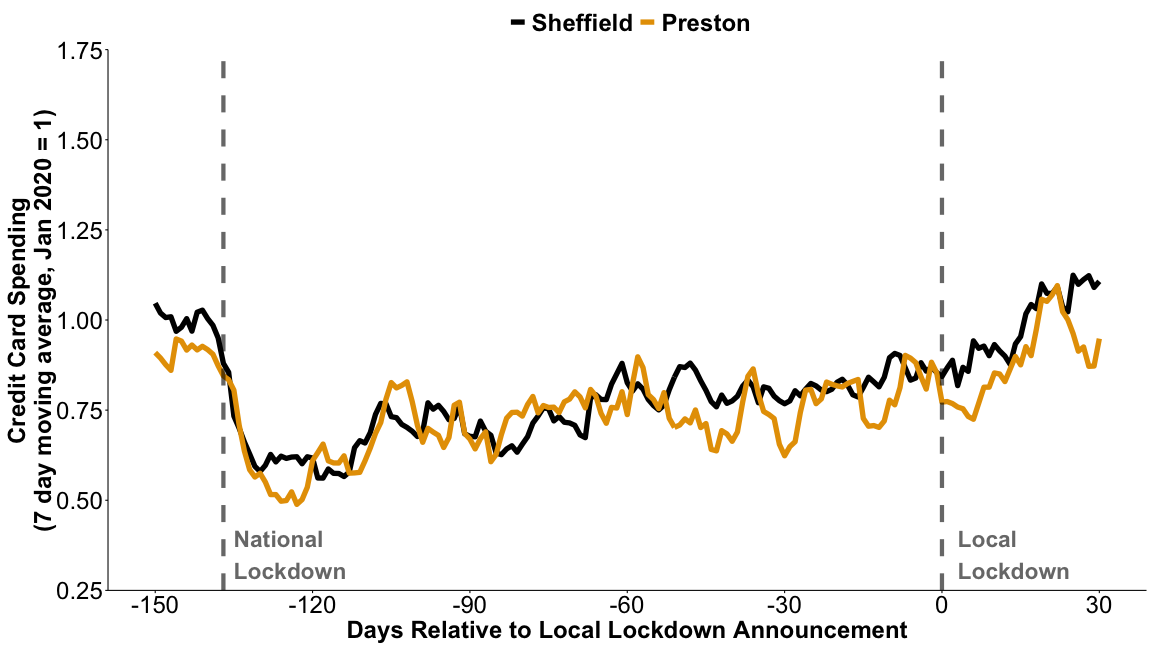}} \\ 
\textbf{J. Leeds} &
\textbf{K. Newcastle} &
\textbf{L. Wolverhampton}
 \\
{\includegraphics[height=1in]{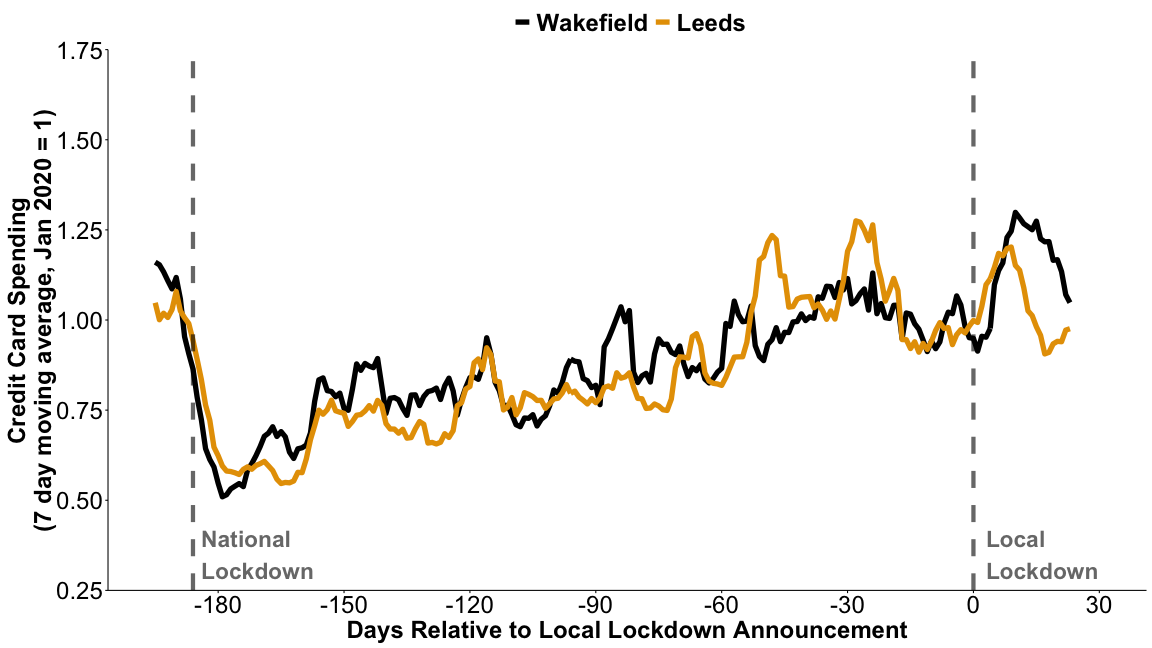}} & {\includegraphics[height=1in]{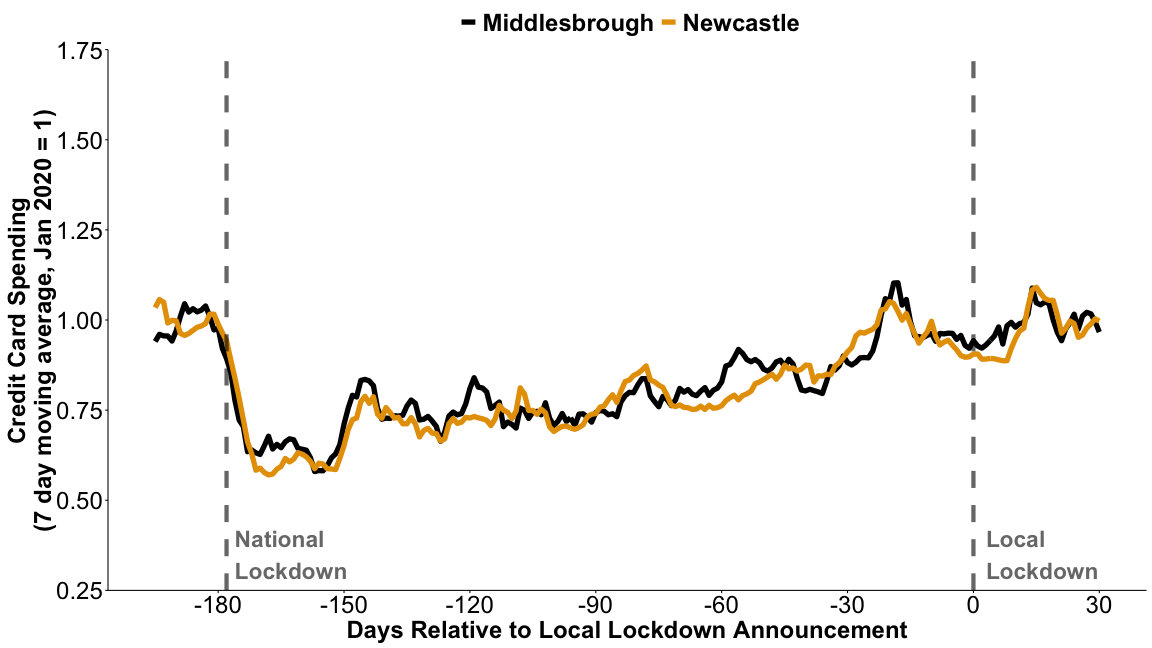}} &
{\includegraphics[height=1in]{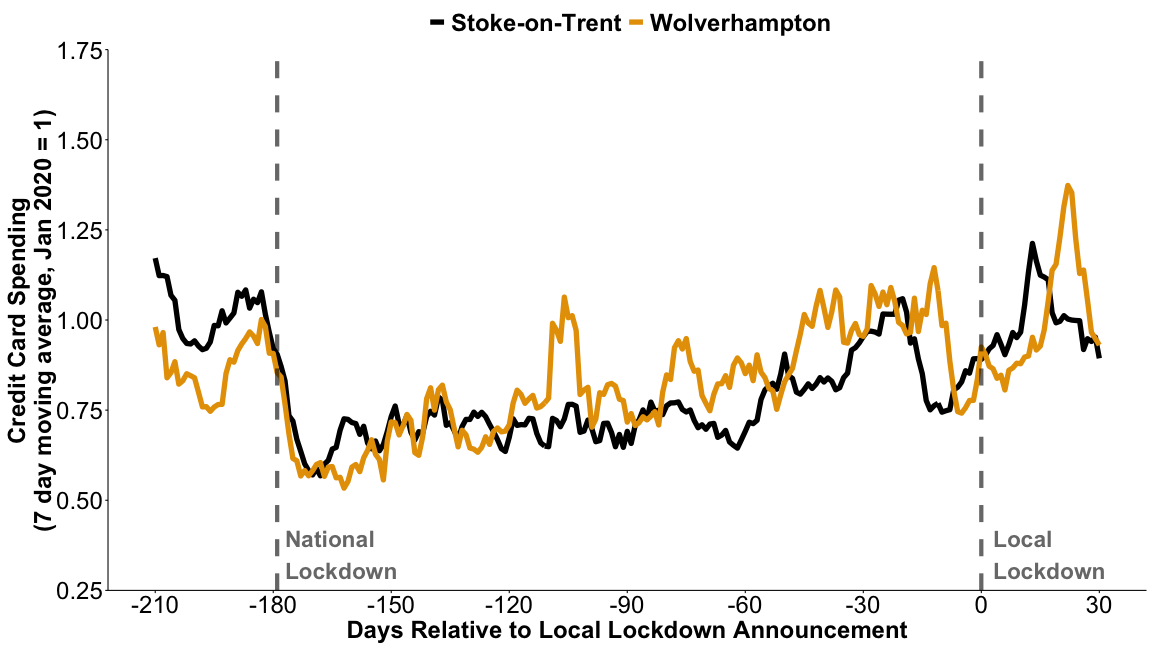}} \\ 
\end{tabular}
\label{fig:allgrid}
\begin{tablenotes}
\small
\item \textit{Notes: Fable Data. Overall credit card spending is a 7 day moving average de-seasoned by taking ratio of the 7 day moving average a year prior. The series is then indexed to its moving average 8 - 28 January 2020.}
\end{tablenotes}
	\label{fig:allgrid}
	\end{figure}

\newpage

\begin{figure}[H]
\caption{\textbf{Food and beverage credit card spending in areas subject to local lockdown (yellow) compared to control areas not locked down (black)}} 
\centering
\begin{tabular}{c c c} \\
\textbf{A. Aberdeen} &
\textbf{B. Birmingham} &
\textbf{C. Bolton} \\
{\includegraphics[height=1in]{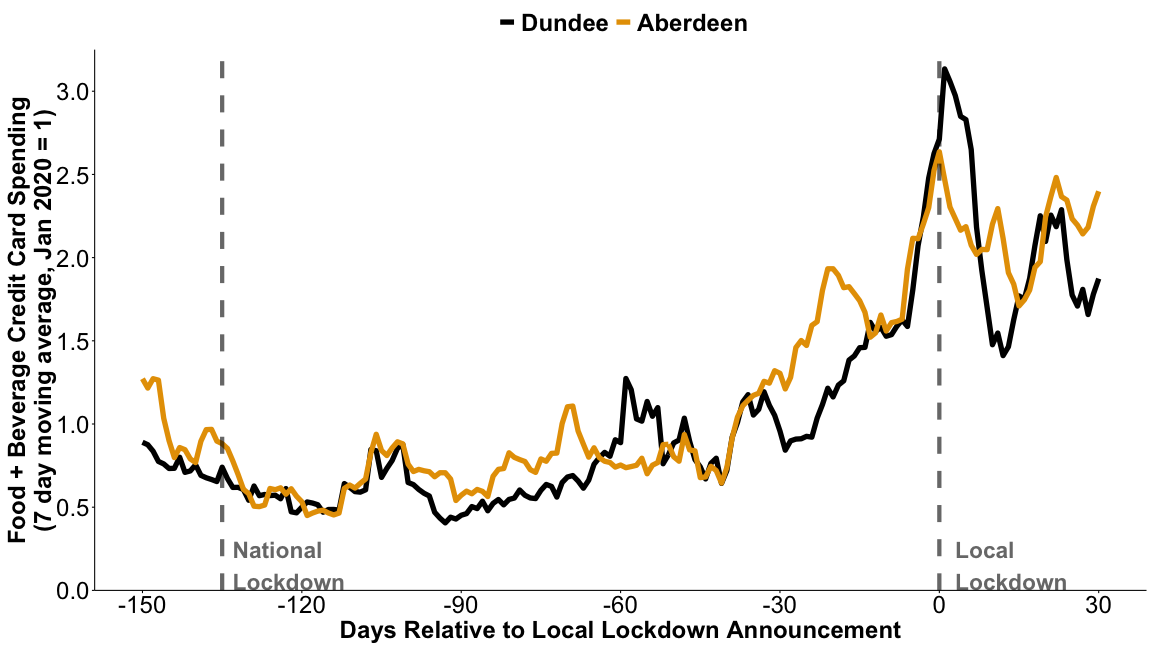}} & {\includegraphics[height=1in]{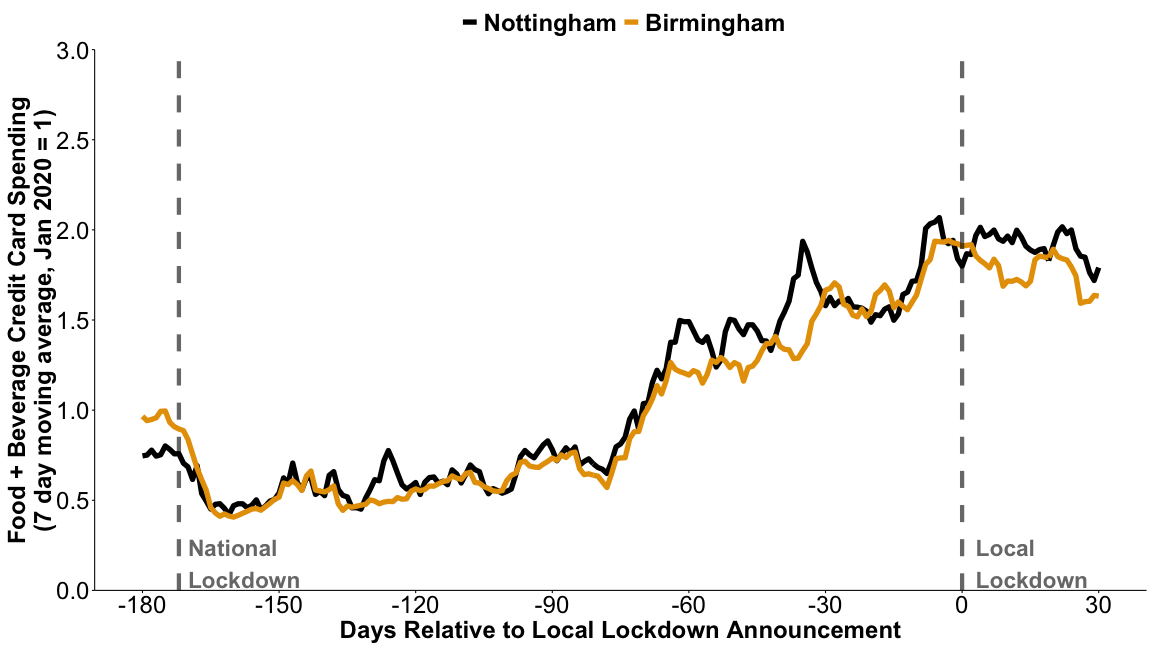}} & {\includegraphics[height=1in]{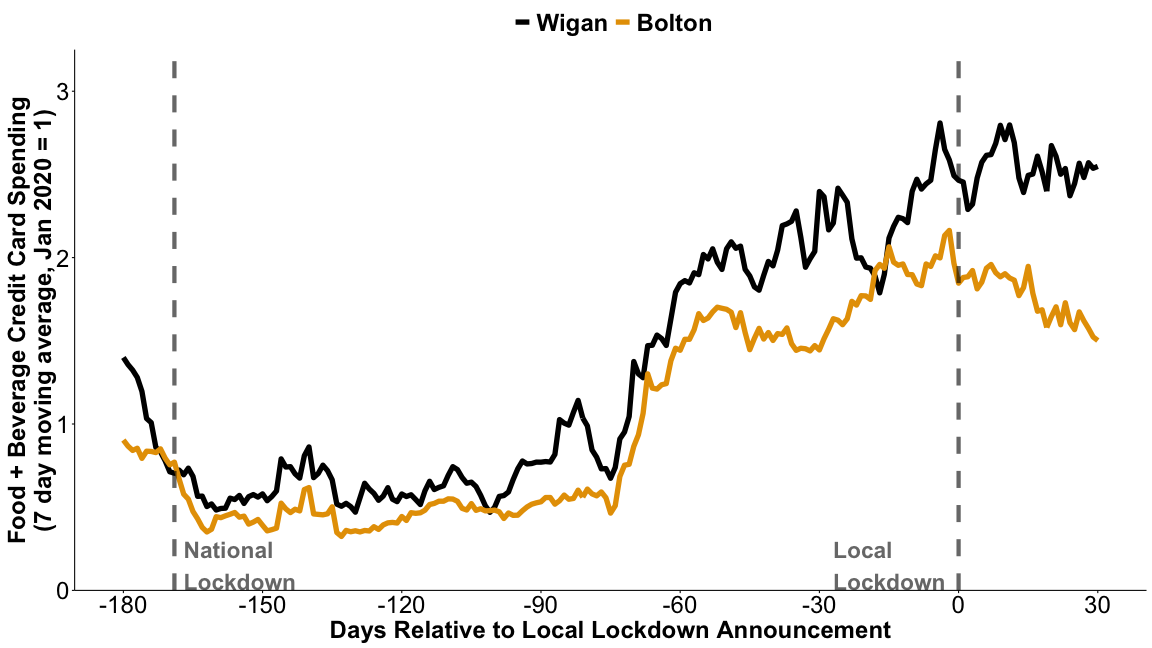}} \\ 
\textbf{D. Caerphilly} &
\textbf{E. Glasgow} &
\textbf{F. Greater Glasgow} \\
{\includegraphics[height=1in]{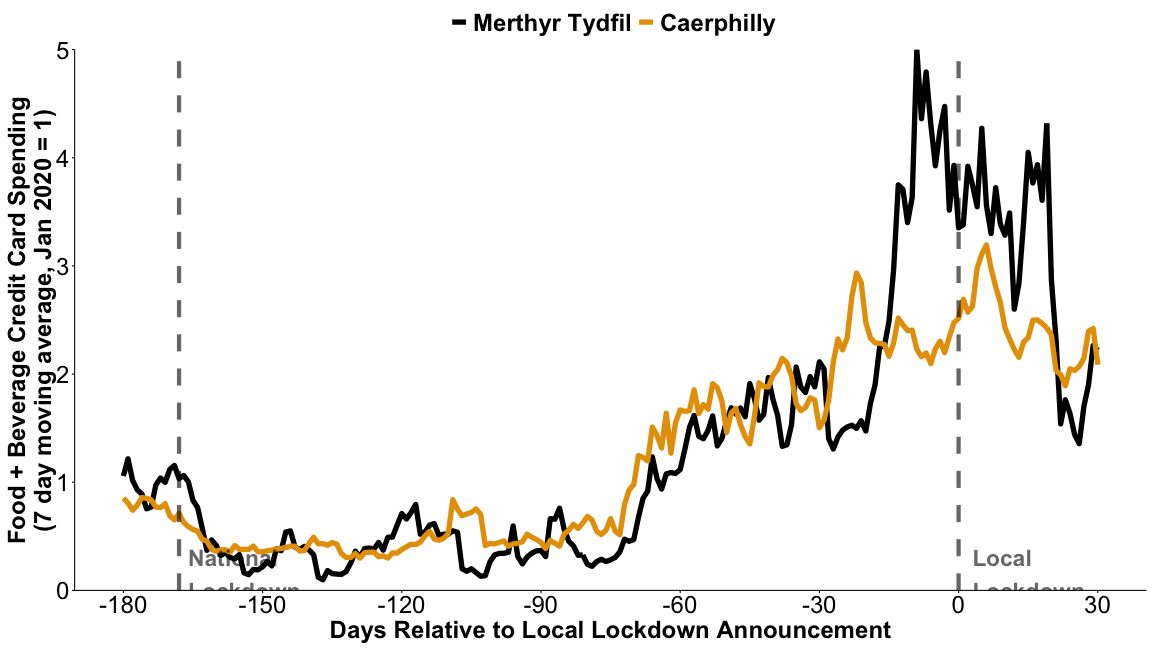}} & {\includegraphics[height=1in]{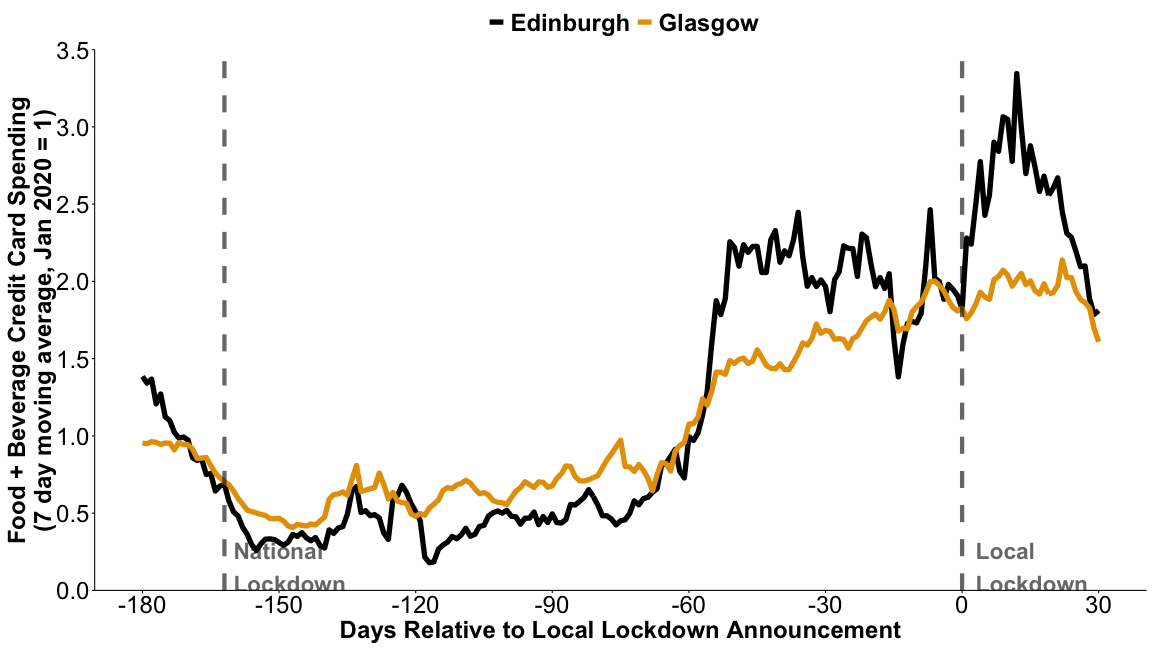}} & {\includegraphics[height=1in]{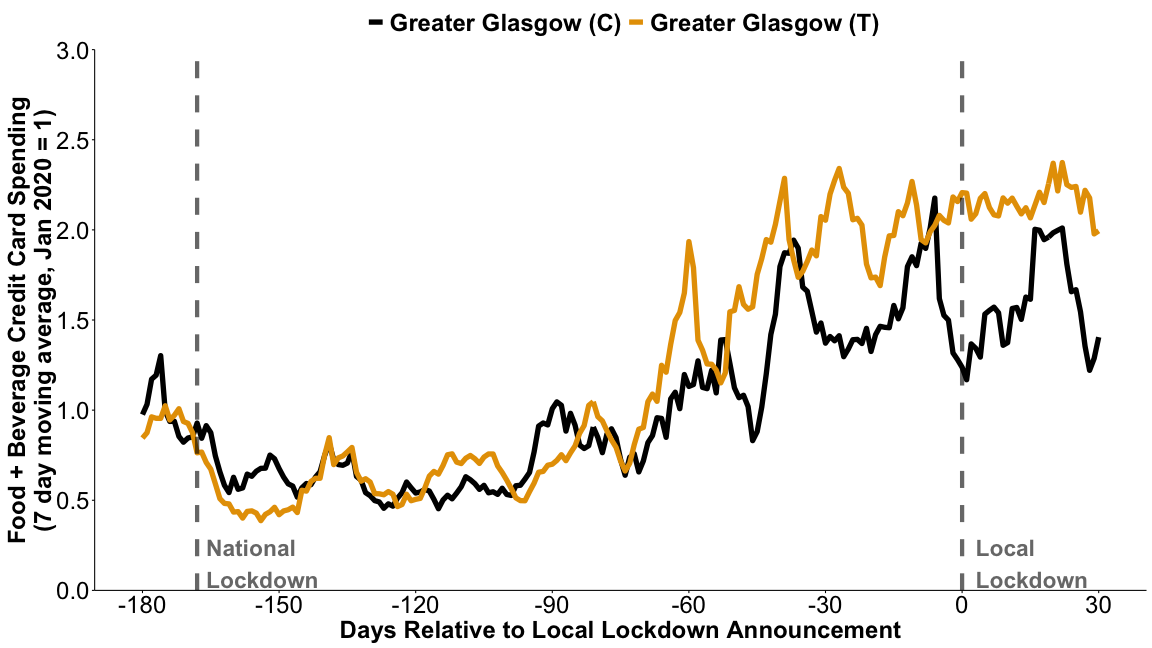}} \\ 
\textbf{G. Lanarkshire} &
\textbf{H. Leicester} &
\textbf{I. Preston} \\
{\includegraphics[height=1in]{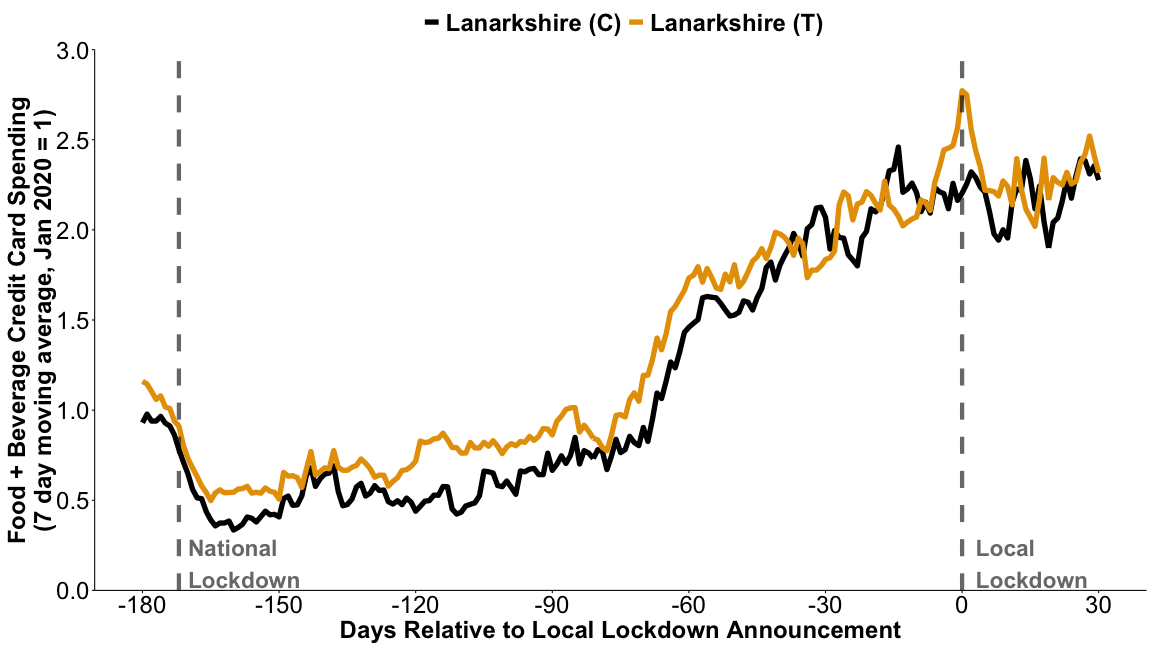}} & {\includegraphics[height=1in]{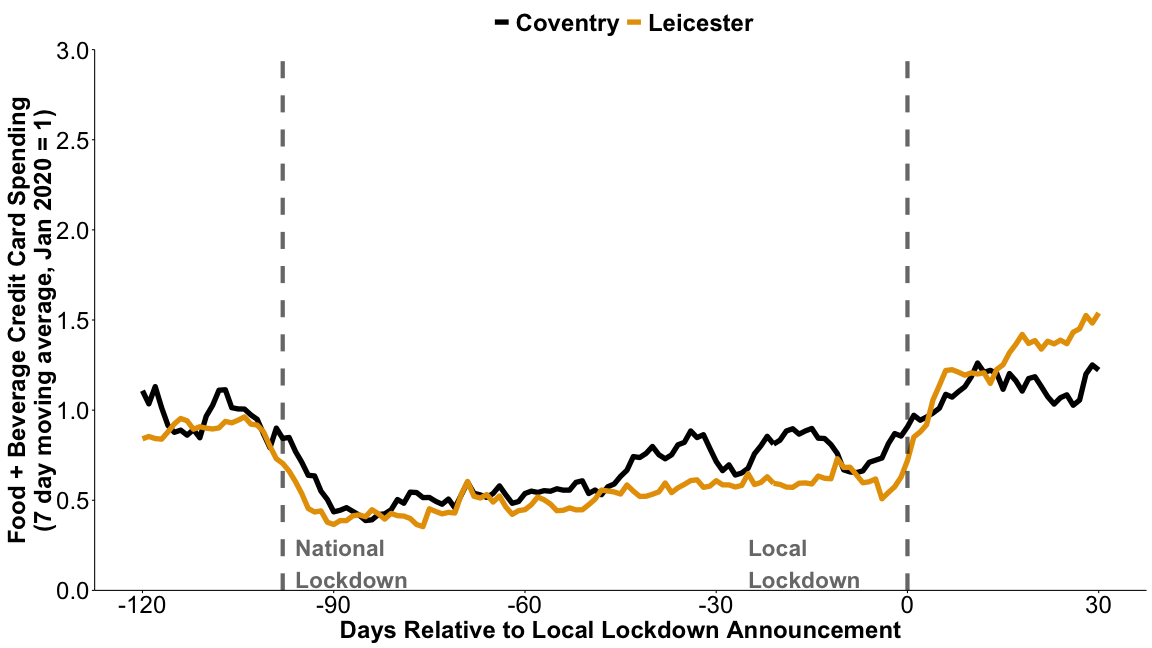}} & {\includegraphics[height=1in]{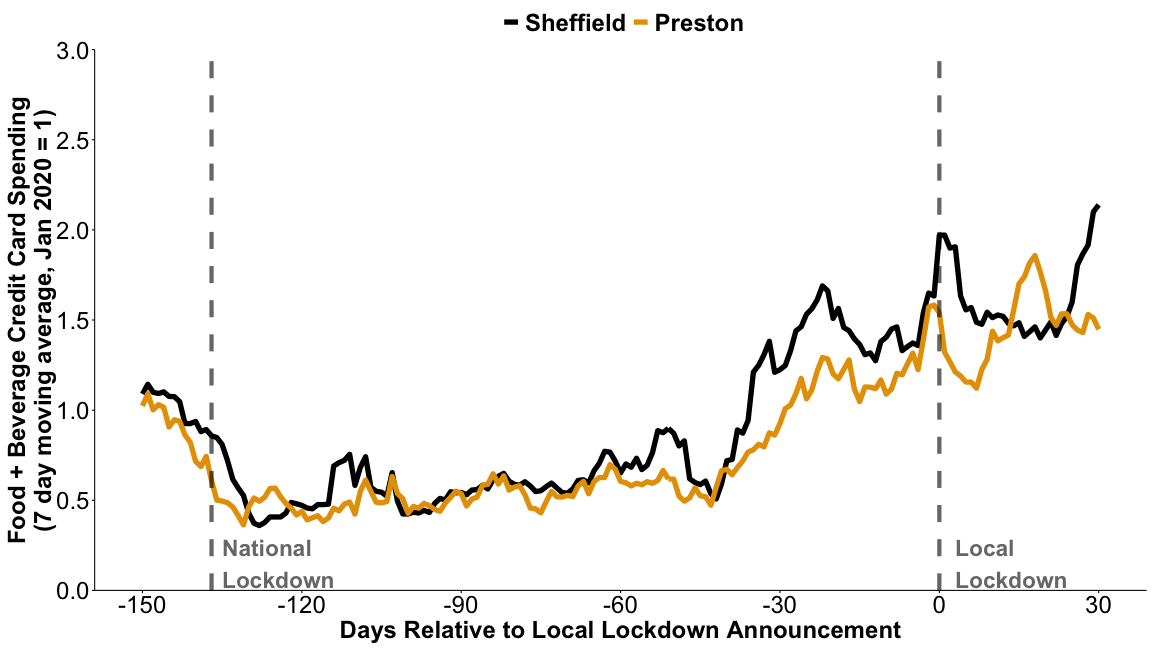}} \\ 
\textbf{J. Leeds} &
\textbf{K. Newcastle} &
\textbf{L. Wolverhampton} \\
{\includegraphics[height=1in]{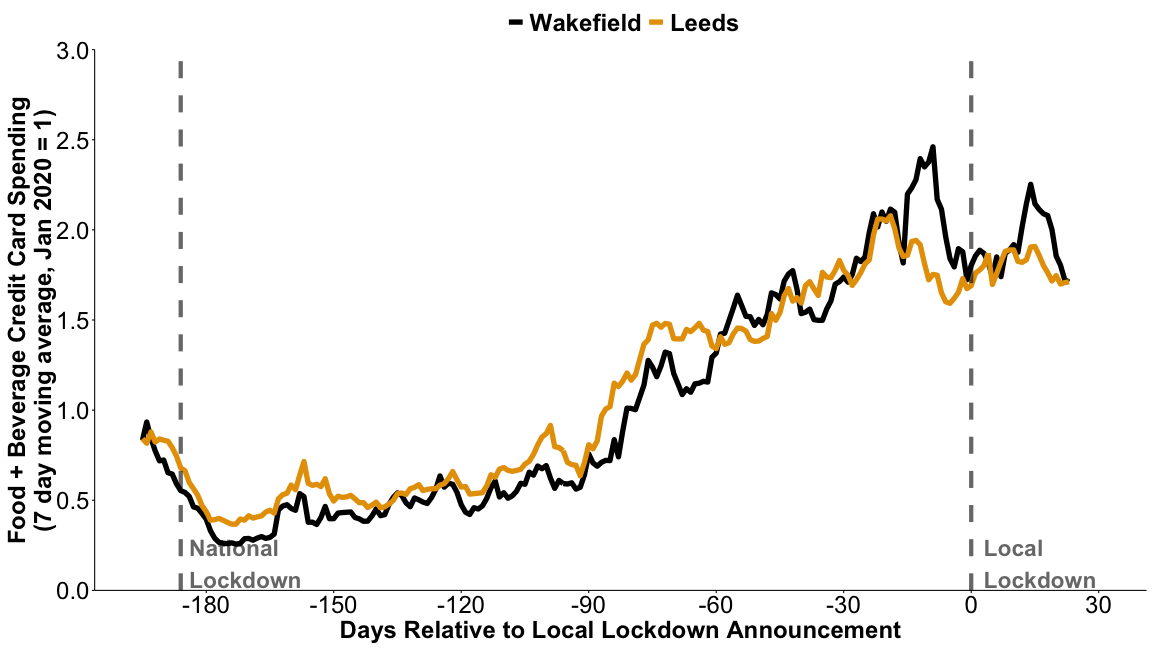}} & {\includegraphics[height=1in]{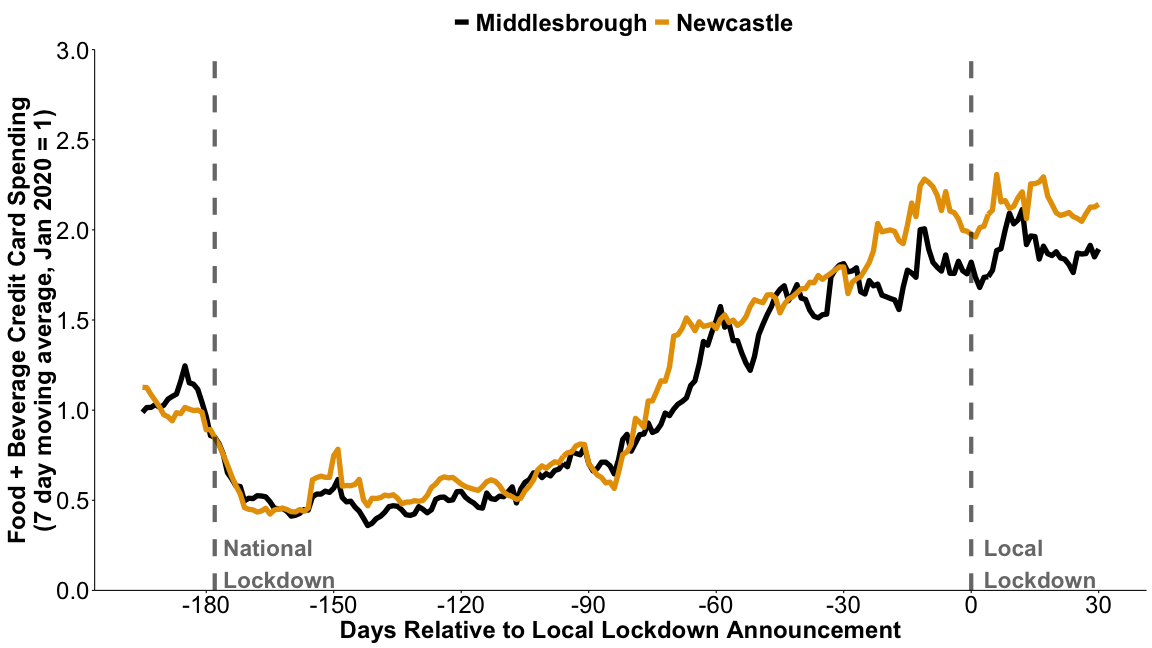}} & {\includegraphics[height=1in]{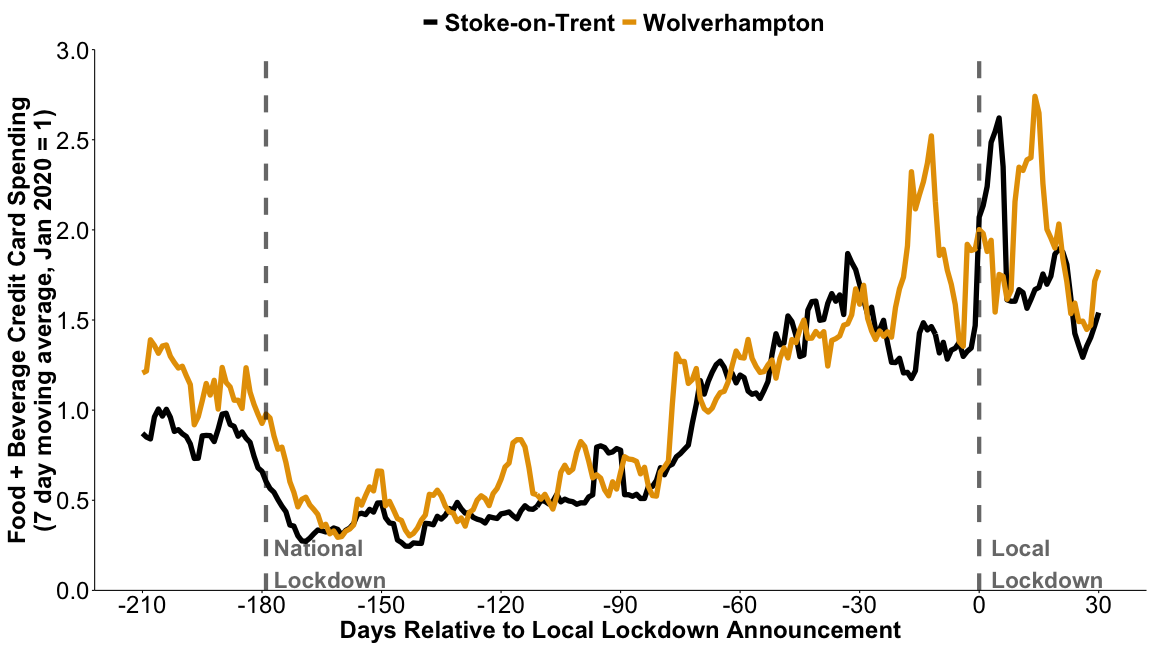}} \\ 
\end{tabular}
\label{fig:fbgrid}
\begin{tablenotes}
\small
\item \textit{Notes: Fable Data. Food and beverage categorization is based on Fable Data categorization using merchant category codes and transaction labels. Credit card spending is a 7 day moving average de-seasoned by taking ratio of the 7 day moving average a year prior. The series is then indexed to its moving average 8 - 28 January 2020.}
\end{tablenotes}
\label{fig:fbgrid}
\end{figure}

\newpage

\begin{figure}[H]
\caption{\textbf{Credit card spending in large store chains areas subject to local lockdown (yellow) compared to control areas not locked down (black), 14 day moving average}} 
\centering
\begin{tabular}{c c c} \\
\textbf{A. Aberdeen} &
\textbf{B. Birmingham} &
\textbf{C. Bolton} \\
{\includegraphics[height=1in]{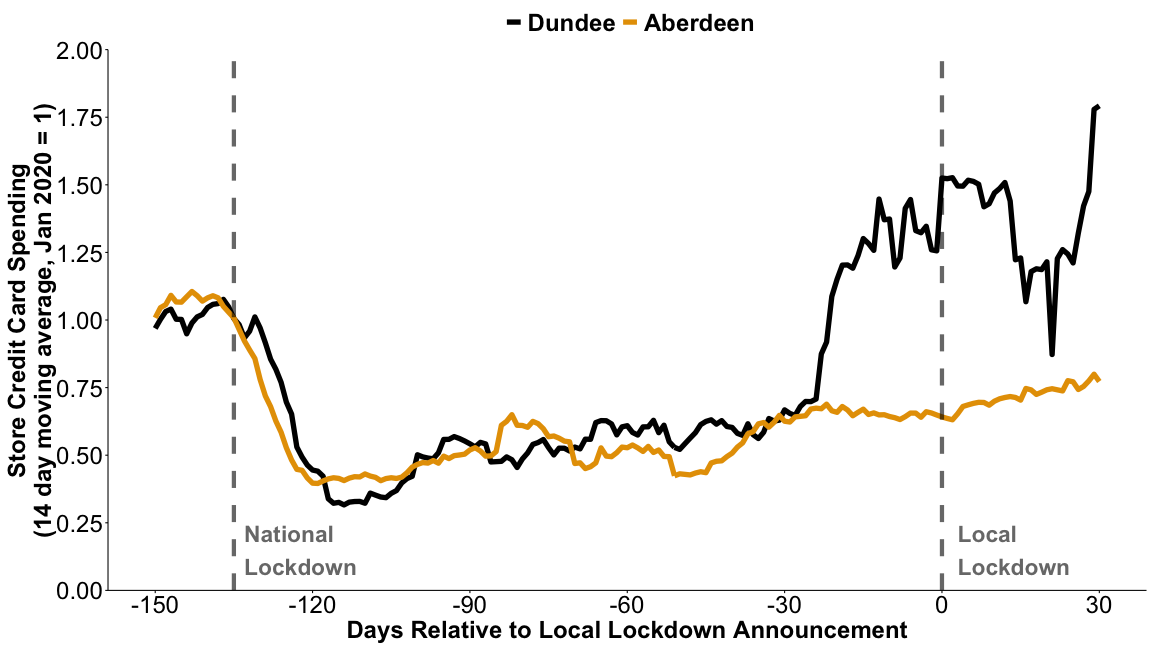}} & {\includegraphics[height=1in]{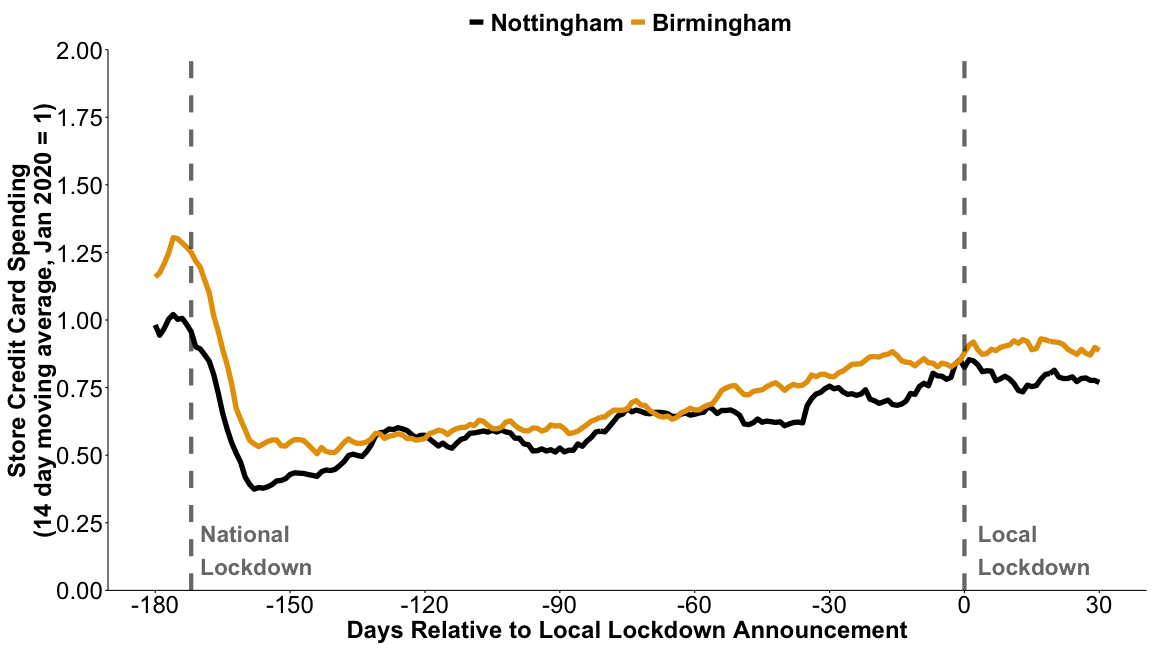}} & {\includegraphics[height=1in]{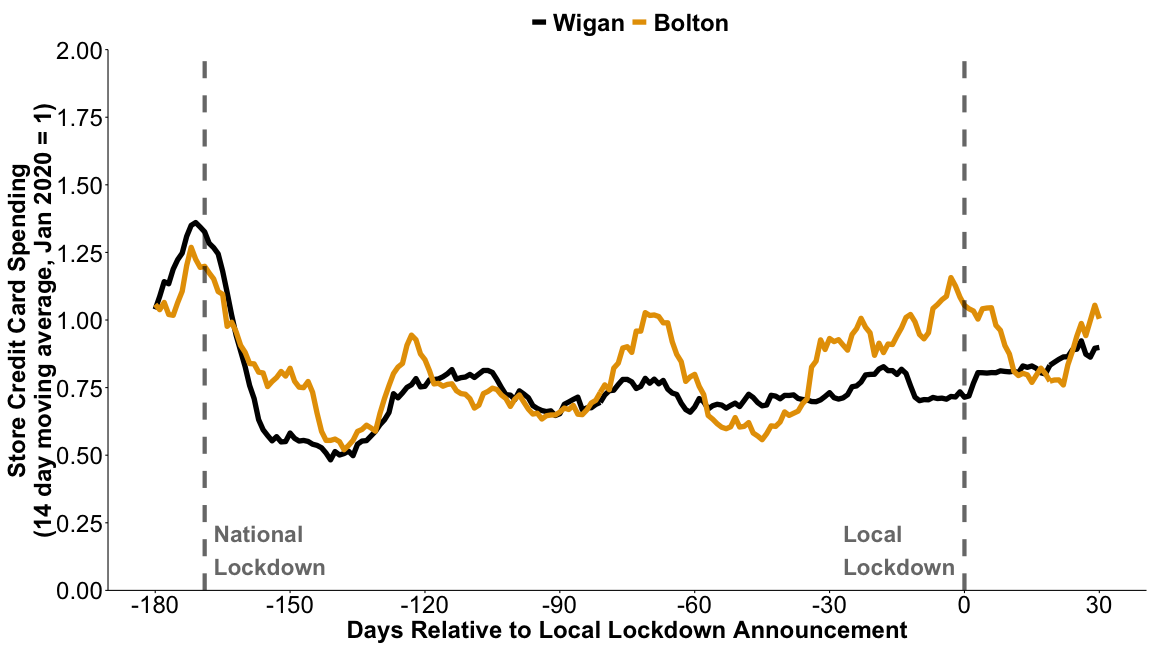}} \\ 
\textbf{D. Caerphilly} &
\textbf{E. Glasgow} &
\textbf{F. Greater Glasgow} \\
{\includegraphics[height=1in]{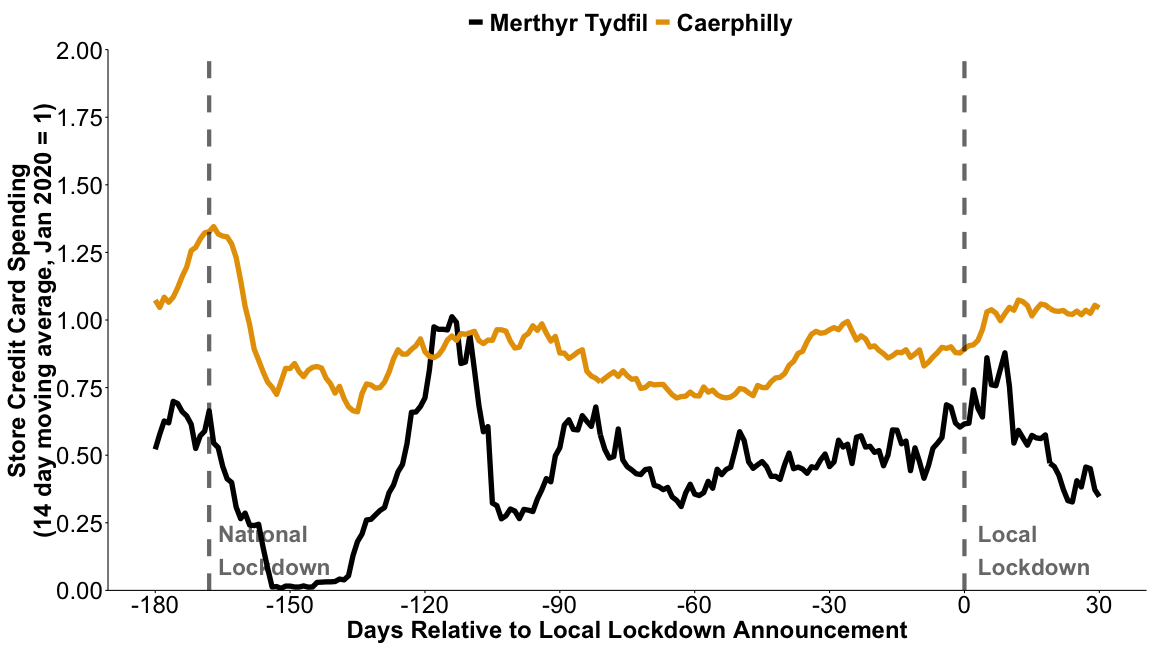}} & {\includegraphics[height=1in]{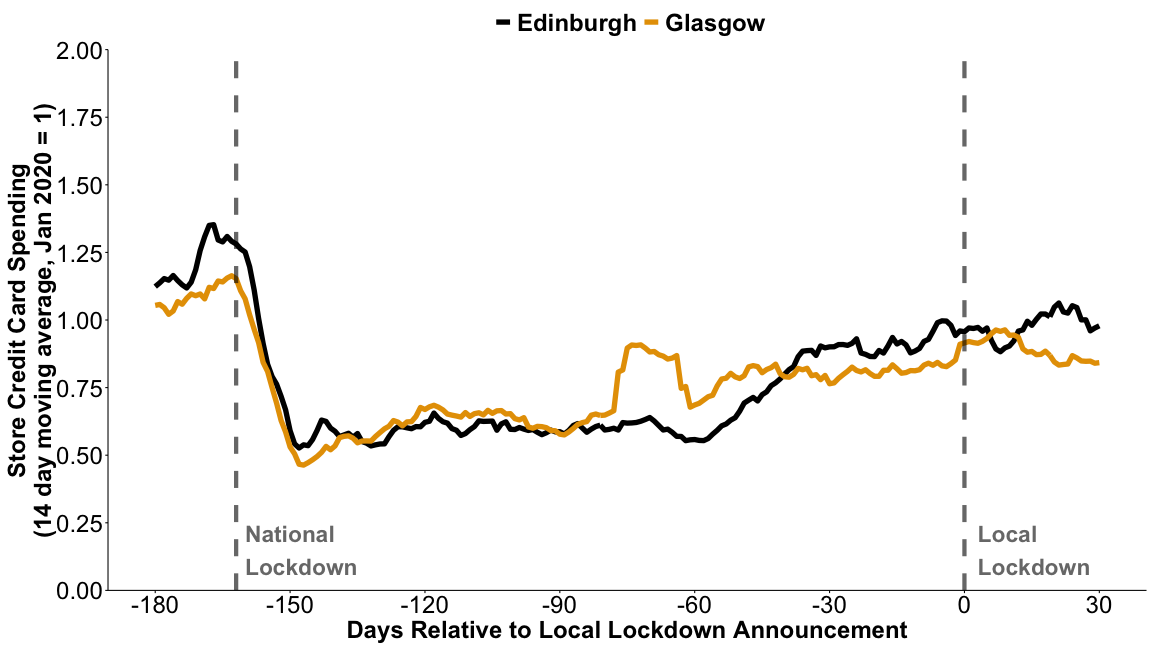}} & {\includegraphics[height=1in]{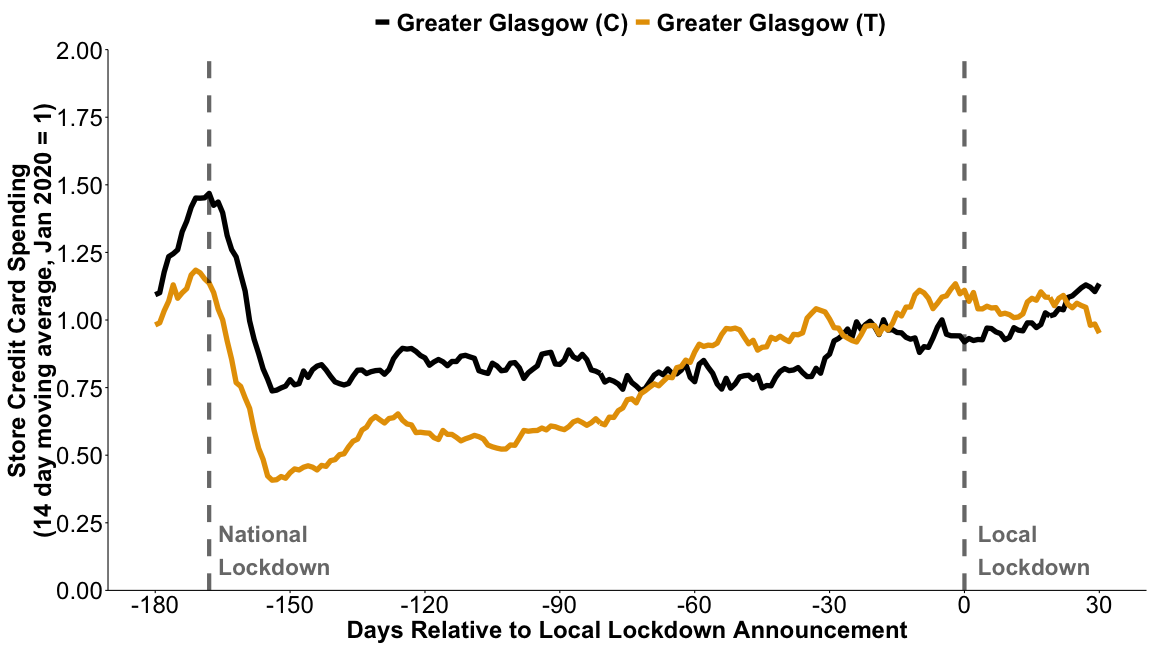}} \\ 
\textbf{G. Lanarkshire} &
\textbf{H. Leicester} &
\textbf{I. Preston} \\
{\includegraphics[height=1in]{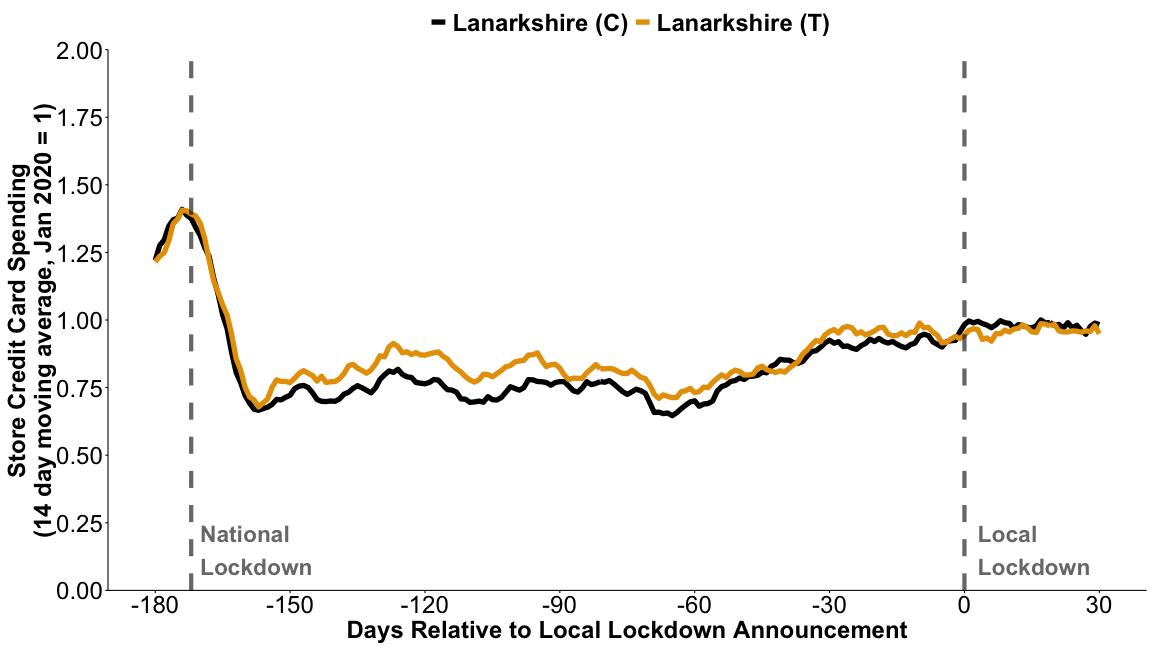}} & {\includegraphics[height=1in]{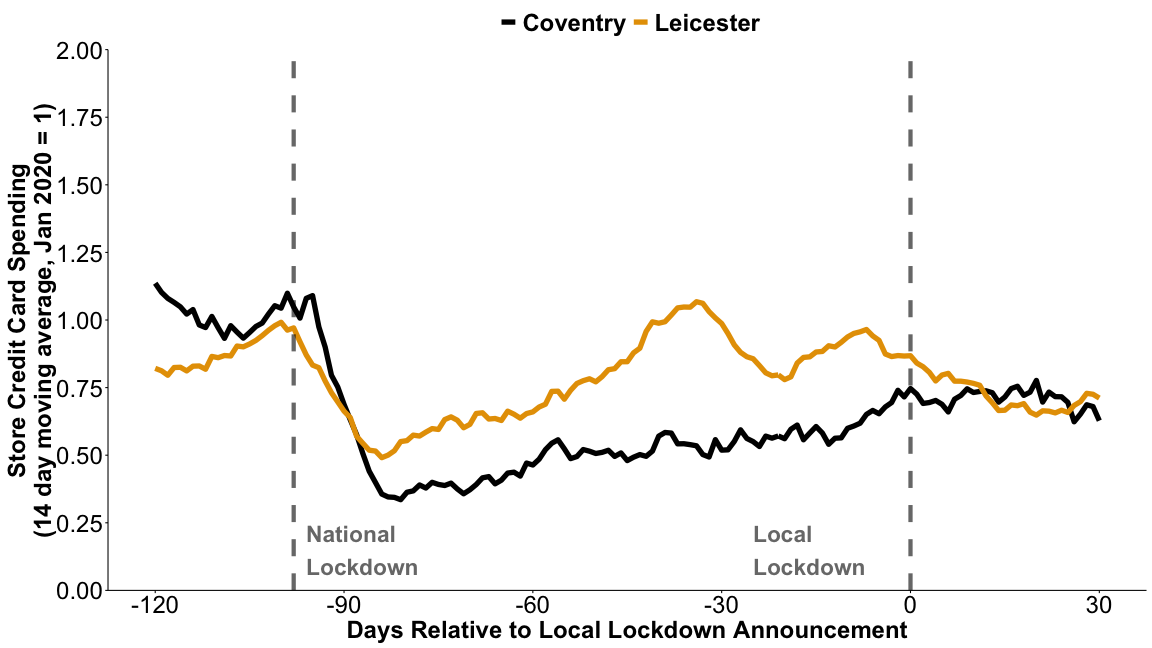}} & {\includegraphics[height=1in]{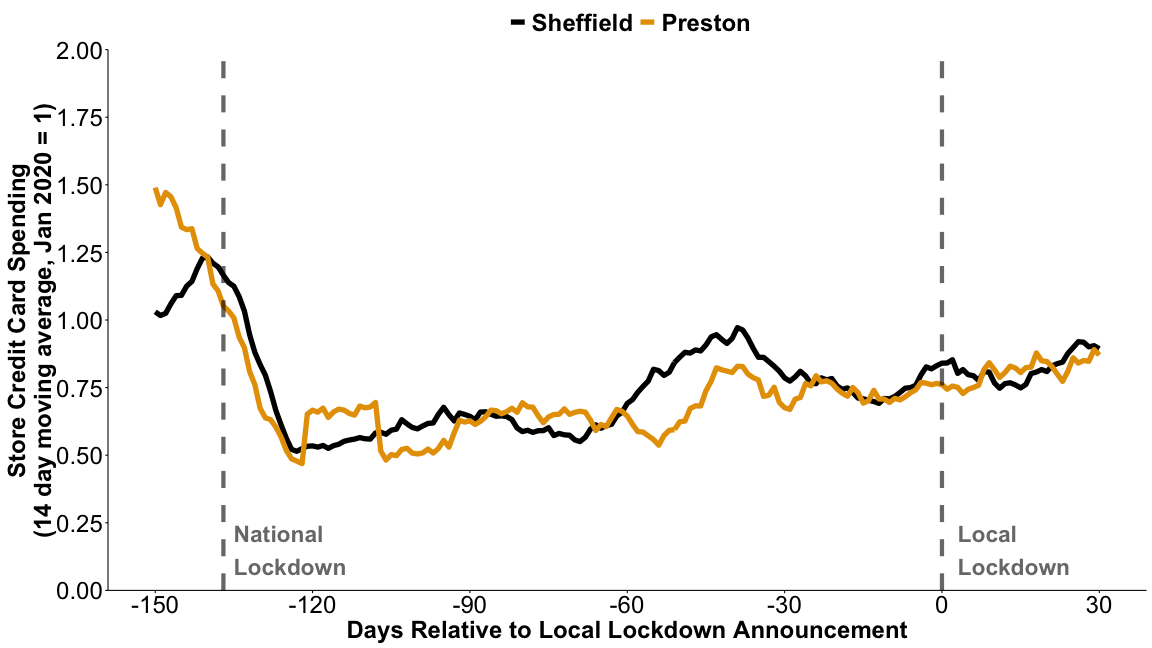}} \\ 
\textbf{J. Leeds} &
\textbf{K. Newcastle} &
\textbf{L. Wolverhampton} \\
{\includegraphics[height=1in]{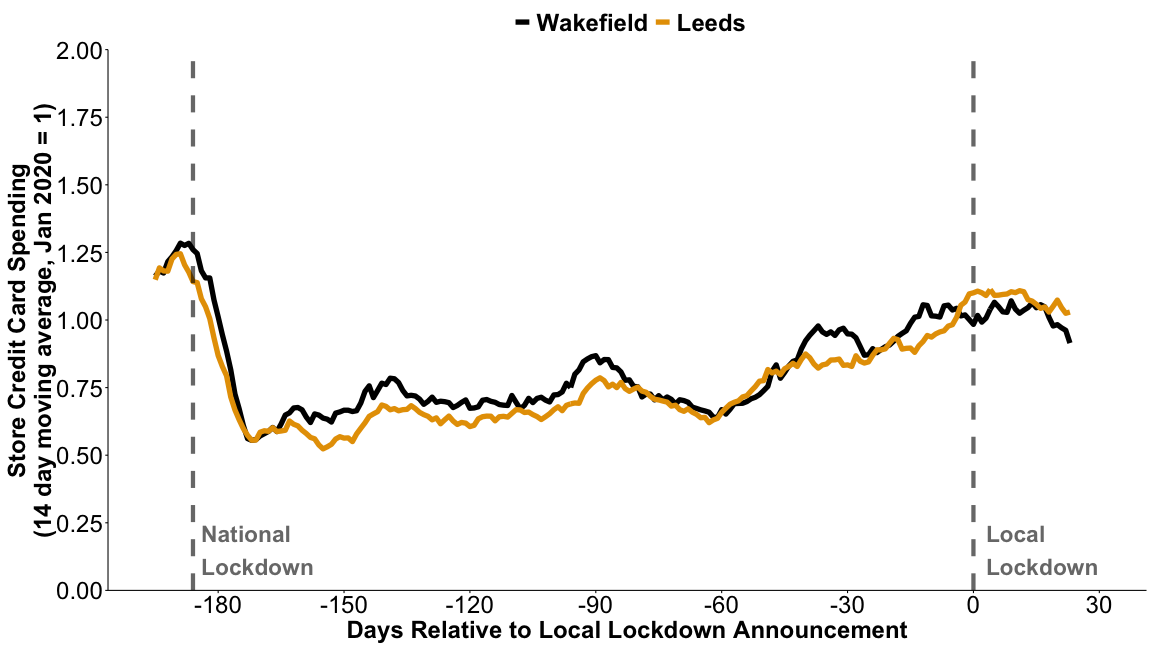}} & {\includegraphics[height=1in]{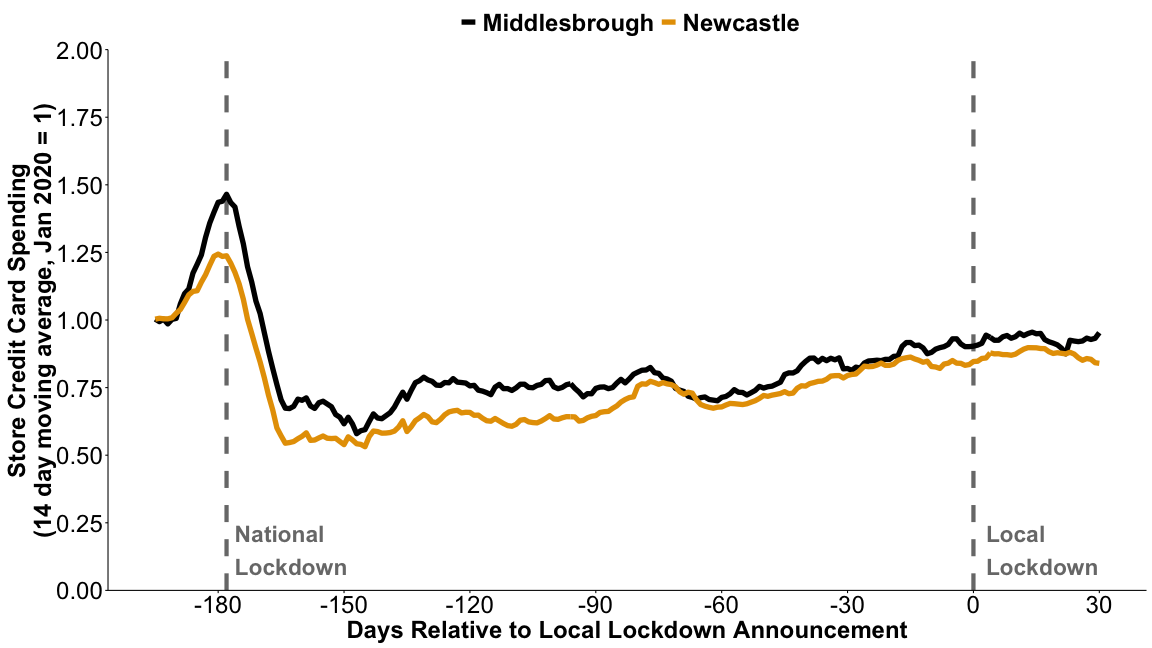}} & {\includegraphics[height=1in]{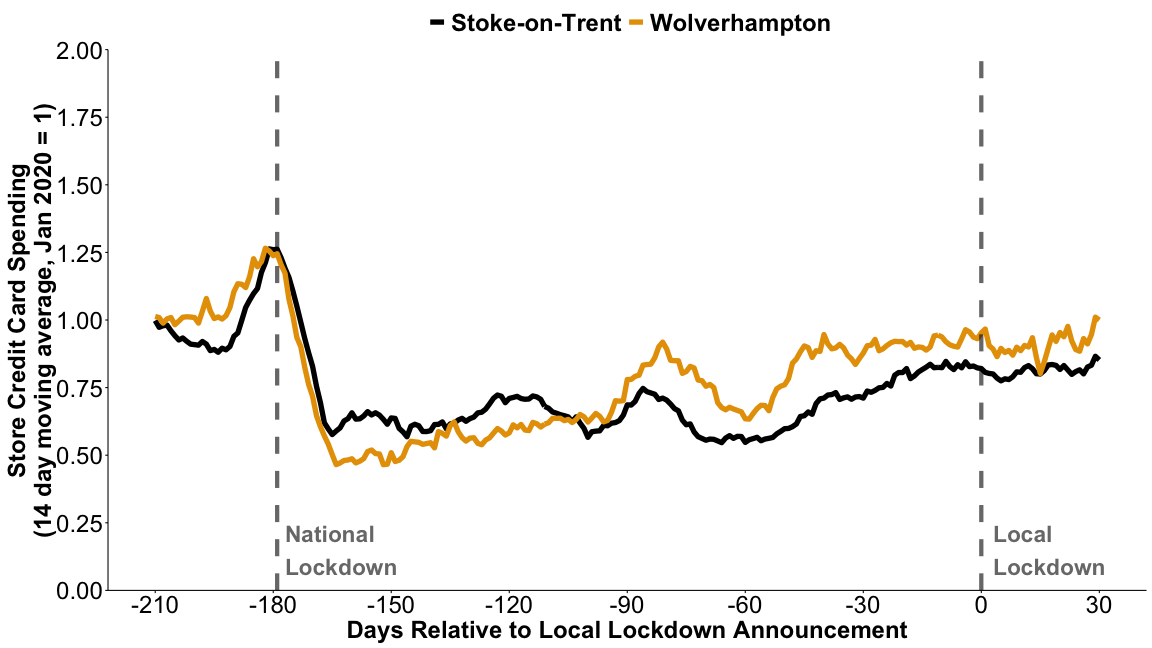}} \\ 
\end{tabular}
\begin{tablenotes}
\small
\item \textit{Notes: Fable Data. Store spending based on transactions tagged to large retail store chain locations. Credit card spending is a 14 day moving average de-seasoned by taking ratio of the 14 day moving average a year prior. The series is then indexed to its moving average 8 - 28 January 2020.}
\end{tablenotes}
\label{fig:storegrid}
\end{figure}


\end{document}